\documentclass[aps,prc,showpacs,preprintnumbers,superscriptaddress,twocolumn,groupedaddress]{revtex4} 
\usepackage{graphicx}
\usepackage{dcolumn}
\usepackage{bm}
\usepackage{amsmath} 
 
\begin{document} 
 
\title{An isospin dependent global nucleon-nucleus optical model \\
           at intermediate energies }

\author{S.~P.~Weppner}
\email{weppnesp@eckerd.edu}
\affiliation{Natural Sciences, Eckerd College, 
             St. Petersburg, FL, 33711}

\author{R. B. Penney\footnote {Present Address:Department of Physics,
Florida State University, Tallahassee, FL, 32306}}
\affiliation{Natural Sciences, Eckerd College,
             St. Petersburg, FL, 33711}

\author{G. W. Diffendale\footnote{Marine Technology Department, 
Rosenstiel School of Marine and Atmospheric Science, University of Miami, 
Miami, FL, 33149}}
\affiliation{Natural Sciences, Eckerd College,
             St. Petersburg, FL, 33711}

\author{G. Vittorini\footnote{Present Address:Department of Physics,
Georgia Institute of Technology, Atlanta, GA, 30332}}
\affiliation{Natural Sciences, Eckerd College,
             St. Petersburg, FL, 33711}

\date{\today} 
 
\begin{abstract} 
A global nucleon-nucleus optical potential for elastic scattering
has been produced which replicates experimental data to high accuracy
and compares well 
with other recently formulated potentials.
The calculation that has been developed describes proton and neutron
scattering from
target nuclei ranging from carbon to nickel and is applicable for 
projectile energies from 30 to 160 MeV. With these ranges it is suitable
for calculations associated with experiments performed by exotic
beam accelerators. The potential has
both real and imaginary isovector asymmetry terms to better describe the dynamics
of chains of isotopes and mirror nuclei. An analysis of the 
validity and strength of the asymmetry term is included with
connections established to other optical potentials and charge-exchange reaction data.
\end{abstract} 
 
 
\maketitle 

\section{Introduction} 
\label{sec:intro}
The fitting of global nucleon-nucleus optical
model potentials (OMP) for elastic scattering has a venerable 
history~\cite{globalK,globalM,globalB,globalJ2,Ca40p80ra,globalV,globalR,Fe56p40ra,globalMR,globalC}.
The global optical potential of this work (WP OMP)
was specifically designed with the
next generation of radioactive beam accelerators in mind.
It attempts to fit a lighter range
of targets ($12\le A\le60$), including those far from stability,
and it is over a
limited projectile energy range (30 MeV $\le E \le 160$ MeV).
This research has produced
one continuous global isospin dependent OMP which incorporates
both proton and neutron scattering
and real and imaginary
isovector asymmetry terms.
Overall 
the potential of this work is recommended 
if one is interested in examining trends in the light and
medium nuclei, specifically isospin asymmetry dependencies, shell closure, 
and neutron excess in isotopes. Since the elastic potential
is often used as a starting point in developing
further inelastic results, this global optical potential will
give the researcher a consistent formulation over a wide
variety of nuclei from which to generate multichannel calculations. To
promote its use we provide an on-line optical potential calculator~\cite{applet}.

Recently there has been the development of two other modern global
OMPs for nucleon-nucleus elastic
scattering which cover an extensive
projectile energy and target range: the potentials of
Koning and Delaroche (KD OMP)~\cite{globalK}
and Madland (MD OMP)~\cite{globalM}. Specifically the potential of
Ref.~{\cite{globalK}}
has set an impressive benchmark for its extensive
projectile energy range (1 keV to 200 MeV)
and its accuracy of fit. We will make theoretical and calculatable comparisons of the
new WP OMP to these noteworthy potentials throughout this article.

In Sec.~\ref{sec:theory} we briefly discuss the theory of the
global optical potential. 
A summary
of our theoretical fitting procedure and the experimental reactions constraining
the fit follows in Sec.~\ref{sec:opt}. A 
generous amount of example calculations, given in Sec.~\ref{sec:results},
compare the results of this potential with the two other contemporary optical 
potentials.
This is followed, in Sec.~\ref{sec:iso}, by a detailed analysis of the 
isovector asymmetric term
identifying dramatic differences in the
magnitude and character of the isovector asymmetry 
$N-Z$ term of the three global OMPs as
well as with standard $t-\rho$ microscopic potentials.
Reactions are isolated which would better 
constrain the isovector term in the future.
We
end with
concluding remarks and future directions.

\section{Theory} 
\label{sec:theory}

The theoretical and computational aspects of creating a global optical
potential has been discussed in detail elsewhere, we only mention a few works
which were found to be especially useful for this research. The seminal work of
M.~A.~Melkanoff et. al.~\cite{melkanoff} 
discusses
in thorough detail many of the hurdles that need to be overcome when
attempting a computational fitting of nuclear scattering data.
F.~D.~Beccetti and G.~W.~Greenless~\cite{globalB} created 
the first comprehensive and still viable global optical potential,
we especially
found poignant their discussion on weighting the various observables.
Recently A.~J.~Koning and J.~P.~Delaroche have provide the nuclear
physics community with an ambitious optical model potential (KD OMP)~\cite{globalK}
as previously mentioned.
The work accompanying the KD OMP was beneficial for its clear 
discussion of theoretical issues and its extensive and comprehensive tables of 
available experimental data.

\subsection{A new potential}
\label{subsec:ours}
Our complex phenemological optical model potential contains the traditional
volume ($V$), surface ($S$), and spin-orbit ($SO$) nuclear terms which are
delineated using the 
standard Woods-Saxon form factors 
\begin{equation}
f_{WS}(r,{\cal R}_i,{\cal A}_i)=(1+\exp\left ((r-{\cal R}_iA^{1/3})/{\cal A}_i)\right)^{-1},
\end{equation}
where ${\cal R}_i$ is the radius parameter 
and ${\cal A}_i$ is the geometric diffusive parameter.
The $i$ is a placeholder index for the $V$ (volume), $S$ (surface) , and 
$SO$ (spin orbit) designations. 
The phenemological optical model potential takes the standard form:
\begin{eqnarray}
{\cal U}(r,E,A,N,Z,{\cal P},MN)=
\quad\quad\quad\quad\quad\quad\quad\quad\quad\quad\quad && \nonumber \\
-{\cal V}_V(E,A,N,Z,{\cal P},MN)f_{WS}(r,{\cal R}_V,{\cal A}_V)&& \nonumber \\
-i{\cal W}_V(E,A,N,Z,{\cal P},MN)
f_{WS}(r,{\cal R}_V,{\cal A}_V)&& \nonumber \\
+4{\cal A}_S{\cal V}_D(E,A) \frac{d}{dr}f_{WS}(r,{\cal R}_S,{\cal A}_S)&&
\nonumber \\
+i4{\cal A}_S{\cal W}_D(E,A,N,Z,{\cal P})
\frac{d}{dr}f_{WS}(r,{\cal R}_S,{\cal A}_S) &&\nonumber \\
+\frac{2}{r}{\cal V}_{SO}(E,A,N,Z,{\cal P})
\frac{d}{dr}f_{WS}(r,{\cal R}_{SO},{\cal A}_{SO})({\bf l\cdot\sigma}) &&\nonumber \\
+i\frac{2}{r}{\cal W}_{SO}(E,A,N,Z,{\cal P})
\frac{d}{dr}f_{WS}(r,{\cal R}_{SO},{\cal A}_{SO})({\bf l\cdot\sigma}) &&\nonumber \\
+f_{coul}(r,{\cal R}_C,A,N,Z,{\cal P}),&&
\label{WS} 
\end{eqnarray}
where the ${\cal V}_i$ and ${\cal W}_i$ are the real and imaginary potential
amplitudes respectively and $f_{coul}$ is the 
coulomb term which has the following traditional format:
\begin{eqnarray}
&&f_{coul}(r,{\cal R}_C,A,N,Z,{\cal P}) = \frac{1+{\cal P}}{2}
\frac{Ze^2}{r},\;\;\;\;\;r\ge {\cal R}_C,\label{coul}
\\
&&f_{coul}(r,{\cal R}_C,A,N,Z,{\cal P}) = \nonumber \\
&&\quad\quad\quad\quad\quad\quad\frac{1+{\cal P}}{2}
\frac{Ze^2}{2{\cal R}_C}
\Big(3-\frac{r^2}{{{\cal R}_C}^2}\Big ), r\le {\cal R}_C. \label{coulomb}
\end{eqnarray}
For a neutron projectile this term is set to zero. 
The
amplitudes, radii, and diffusive parameters
have the following dependent variables: \\
$\bullet$ {\bf E} -- projectile nucleon laboratory energy in MeV \\
$\bullet$ {\bf A} -- Atomic number of the target nucleus \\
$\bullet$ {\bf N} -- Number of neutrons in the target nucleus \\
$\bullet$ {\bf Z} -- Number of protons in the target nucleus \\
$\bullet$ {$\cal P$} -- +1 if projectile is a proton, -1 if a neutron \\
$\bullet$ {\bf MN} -- set to 1 if the target is traditionally singly magic \\
\hspace*{12pt} -- set to 2 if the target is traditionally doubly magic \\
\hspace*{12pt} -- otherwise set to 0.

Explicitly the thirteen
Woods-Saxon terms of this OMP
are given using twenty-three distinct parameters where one
Woods-Saxon potential term may have up to three of these parameters. The
systematic polynomial formats of these terms are described below. 
The parameters are
purposely labeled to emphasize that each 
was fitted independently in
the process of arriving at the final WP OMP, the techniques used will be detailed in
Sec.~\ref{sec:opt}.

\begin{widetext}
Here are the polynomial forms of this optical potential
beginning with the volume amplitudes:
\begin{eqnarray}
&&{\cal V}_V=V_{V_0}+V_{V_1}A+V_{V_2}A^2+V_{V_3}A^3
+V_{V_5}E+V_{V_6}E^2+V_{V_7}E^3 \label{start}\\
+ &&{\cal P} (N-Z)\Big(V_{V_{i0}}+V_{V_{i1}}A+V_{V_{i2}}A^2+V_{V_{i3}}A^3+V_{V_{i4}}A^4
+V_{V_{i5}}E+V_{V_{i6}}E^2\Big) \label{eq2}\\
+ &&MN\Big(V_{V_{m0}}+V_{V_{m1}}A+V_{V_{m2}}A^2+V_{V_{m3}}A^3
+V_{V_{m5}}E+V_{V_{m6}}E^2\Big),\label{eq3} 
\end{eqnarray}
\begin{eqnarray}
&&{\cal W}_V=W_{V_0}+W_{V_1}A+W_{V_2}A^2+W_{V_3}A^3
+W_{V_5}E+W_{V_6}E^2+W_{V_7}E^3 \label{eq4}\\
+ &&{\cal P} (N-Z)\Big(W_{V_{i0}}+W_{V_{i1}}A+W_{V_{i2}}A^2+W_{V_{i3}}A^3+W_{V_{i4}}A^4
+W_{V_{i5}}E+W_{V_{i6}}E^2\Big) \label{eq5}\\
+&&MN\Big(W_{V_{m0}}+W_{V_{m1}}A+W_{V_{m2}}A^2+W_{V_{m3}}A^3
+W_{V_{m5}}E+W_{V_{m6}}E^2\Big). \label{eq6}
\end{eqnarray}
The surface amplitudes:
\begin{eqnarray}
&&{\cal V}_S=V_{S_0}+V_{S_1}A+V_{S_2}A^2+V_{S_3}A^3
+V_{S_5}E+V_{S_6}E^2+V_{S_7}E^3, \label{eq7}
\end{eqnarray}
\begin{eqnarray}
&&{\cal W}_S=W_{S_0}+W_{S_1}A+W_{S_2}A^2+W_{S_3}A^3
+W_{S_5}E+W_{S_6}E^2+W_{S_7}E^3 \label{eq8}\\
+ &&{\cal P} (N-Z)\Big(W_{S_{i0}}+W_{S_{i1}}A+W_{S_{i2}}A^2+W_{S_{i3}}A^3+W_{S_{i4}}A^4
+W_{S_{i5}}E+W_{S_{i6}}E^2\Big). \label{eq9}
\end{eqnarray}
The spin orbit amplitudes:
\begin{eqnarray}
&&{\cal V}_{SO}=V_{{SO}_0}+V_{{SO}_1}A+V_{{SO}_2}A^2+V_{{SO}_3}A^3
+V_{{SO}_5}E+V_{{SO}_6}E^2+V_{{SO}_7}E^3 \label{eq10}\\
+ &&{\cal P} (N-Z)\Big(V_{{SO}_{i0}}
+V_{{SO}_{i1}}A+V_{{SO}_{i2}}A^2+V_{{SO}_{i3}}A^3+V_{{SO}_{i4}}A^4
+V_{{SO}_{i5}}E+V_{{SO}_{i6}}E^2\Big), \label{eq11}
\end{eqnarray}
\begin{eqnarray}
&&{\cal W}_{SO}=W_{{SO}_0}+W_{{SO}_1}A+W_{{SO}_2}A^2+W_{{SO}_3}A^3
+W_{{SO}_5}E+W_{{SO}_6}E^2+W_{{SO}_7}E^3 \label{eq12}\\
+&&{\cal P} (N-Z)\Big(W_{{SO}_{i0}}+W_{{SO}_{i1}}A
+W_{{SO}_{i2}}A^2+W_{{SO}_{i3}}A^3+W_{{SO}_{i4}}A^4
+W_{{SO}_{i5}}E+W_{{SO}_{i6}}E^2\Big). \label{eq13}
\end{eqnarray}
The volume radius and diffusive terms:
\begin{eqnarray}
&&{\cal R}_V=R_{V_0}+R_{V_1}A+R_{V_2}A^2+R_{V_3}A^3
+R_{V_5}E+R_{V_6}E^2+R_{V_7}E^3, \label{eq14}
\end{eqnarray}
\begin{eqnarray}
&&{\cal A}_V=A_{V_0}+A_{V_1}A+A_{V_2}A^2+A_{V_3}A^3
+A_{V_5}E+A_{V_6}E^2+A_{V_7}E^3 \label{eq15}\\
+&&{\cal P}(N-Z)\Big(A_{V_{i0}}+A_{V_{i1}}A+A_{V_{i2}}A^2+A_{V_{i3}}A^3
+A_{V_{i5}}E+A_{V_{i6}}E^2+A_{V_{i7}}E^3\Big).\label{eq16}
\end{eqnarray}
The surface radius and diffusive terms:
\begin{eqnarray}
&&{\cal R}_S=R_{S_0}+R_{S_1}A+R_{S_2}A^2+R_{S_3}A^3
+R_{S_5}E+R_{S_6}E^2+R_{S_7}E^3, \label{eq17}
\end{eqnarray}
\begin{eqnarray}
&&{\cal A}_S=A_{S_0}+A_{S_1}A+A_{S_2}A^2+A_{S_3}A^3
+A_{S_5}E+A_{S_6}E^2+A_{S_7}E^3, \label{eq18}
\end{eqnarray}
the spin-orbit radius and diffusive terms:
\begin{eqnarray}
&&{\cal R}_{SO}=R_{{SO}_0}+R_{{SO}_1}A+R_{{SO}_2}A^2+R_{{SO}_3}A^3
+R_{{SO}_5}E+R_{{SO}_6}E^2+R_{{SO}_7}E^3 \label{eq19}\\
+&&{\cal P}(N-Z)\Big(R_{{SO}_{i0}}+R_{{SO}_{i1}}A+R_{{SO}_{i2}}A^2+R_{{SO}_{i3}}A^3
+R_{{SO}_{i5}}E+R_{{SO}_{i6}}E^2+R_{{SO}_{i7}}E^3\Big), \label{eq20}
\end{eqnarray}
\begin{eqnarray}
&&{\cal A}_{SO}=A_{{SO}_0}+A_{{SO}_1}A+A_{{SO}_2}A^2+A_{{SO}_3}A^3
+A_{{SO}_5}E+A_{{SO}_6}E^2+A_{{SO}_7}E^3, \label{eq21}
\end{eqnarray}
and finally the the coulomb radius term:
\begin{eqnarray}
&&{\cal R}_{C}=R_{{C}_0}+R_{{C}_1}A+R_{{C}_2}A^2+R_{{C}_3}A^3
+R_{{C}_5}E+R_{{C}_6}E^2+R_{{C}_7}E^3 \label{eq22} \\
+&&{\cal P}(N-Z)\Big (\frac{2Z}{A}\Big )^{\frac{1}{3}}
\Big (R_{{C}_{i0}}+R_{{C}_{i1}}A+R_{{C}_{i2}}A^2+R_{{C}_{i3}}A^3+R_{{C}_{i4}}A^4
+R_{{C}_{i5}}E+R_{{C}_{i6}}E^2+R_{{C}_{i7}}E^3\Big ).
\label{finish}
\end{eqnarray}
\end{widetext}
\begin{table*}
\begin{tabular}{|l||c|c|c|c|c|c|c|c|}
 \hline
 Term & $0$  & $1\;(A)$ & $2\;(A^2)$ & $3\;(A^3)$
       & $4\;(A^4)$ & $5\;(E)$ & $6\;(E^2)$ & $7\; (E^3)$\\
 \hline
$V$&+5.703$\times 10^{+1}$&+4.099$\times 10^{-1}$&-8.656$\times 10^{-3}$&
+5.793$\times 10^{-5}$&---&-5.881$\times 10^{-1}$&+1.822$\times 10^{-3}$&---\\\hline
$V_i$&-7.810$\times 10^{+0}$&+1.054$\times 10^{+0}$&-4.616$\times 10^{-2}$&
+8.384$\times 10^{-4}$&-5.416$\times 10^{-6}$&-6.729$\times 10^{-3}$&+3.684$\times 10^{-5}$&---\\\hline
$V_m$&-3.723$\times 10^{-1}$&+6.563$\times 10^{-3}$&-5.308$\times 10^{-4}$&
+7.987$\times 10^{-6}$&---&+2.515$\times 10^{-3}$&-5.607$\times 10^{-6}$&---\\\hline
$W$&-1.897$\times 10^{+0}$&-1.843$\times 10^{-1}$&+5.034$\times 10^{-3}$&
-3.814$\times 10^{-5}$&---&+2.367$\times 10^{-1}$&-1.423$\times 10^{-3}$&2.556$\times 10^{-6}$\\\hline
$W_i$&+8.216$\times 10^{+0}$&-8.359$\times 10^{-1}$&+3.221$\times 10^{-2}$&
-5.426$\times 10^{-4}$&+3.320$\times 10^{-6}$&+8.446$\times 10^{-3}$&-2.644$\times 10^{-5}$&---\\\hline
$W_m$&-3.781$\times 10^{+0}$&+1.818$\times 10^{-1}$&-4.772$\times 10^{-3}$&
+3.374$\times 10^{-5}$&---&+4.157$\times 10^{-2}$&-2.149$\times 10^{-4}$&---\\\hline
$V_S$&-4.612$\times 10^{-1}$&-1.178$\times 10^{-2}$&+9.658$\times 10^{-4}$&
-1.270$\times 10^{-5}$&---&+7.906$\times 10^{-3}$&-4.230$\times 10^{-5}$&---\\\hline
$W_S$&+6.189$\times 10^{+0}$&+1.740$\times 10^{-1}$&-4.790$\times 10^{-3}$&
+3.670$\times 10^{-5}$&---&-6.423$\times 10^{-2}$&-3.753$\times 10^{-4}$&+3.096$\times 10^{-6}$\\\hline
$W_{S_i}$&+3.471$\times 10^{+0}$&-4.265$\times 10^{-1}$&+1.670$\times 10^{-2}$&
-2.828$\times 10^{-4}$&+1.744$\times 10^{-6}$&+1.449$\times 10^{-2}$&-8.093$\times 10^{-5}$&---\\\hline
$V_{SO}$&+1.562$\times 10^{+1}$&-1.202$\times 10^{-1}$&+1.765$\times 10^{-3}$&
---&---&-1.923$\times 10^{-1}$&+1.168$\times 10^{-3}$&-2.400$\times 10^{-6}$\\\hline
$V_{SO_i}$&-3.666$\times 10^{+0}$&+7.228$\times 10^{-1}$&-3.524$\times 10^{-2}$&
+6.493$\times 10^{-4}$&-4.151$\times 10^{-6}$&+2.472$\times 10^{-3}$&-3.317$\times 10^{-6}$&---\\\hline
$W_{SO}$&+3.929$\times 10^{-1}$&+1.660$\times 10^{-1}$&-5.369$\times 10^{-3}$&
+4.646$\times 10^{-5}$&---&-3.702$\times 10^{-2}$&+9.223$\times 10^{-5}$&---\\\hline
$W_{SO_i}$&+5.399$\times 10^{+0}$&-4.639$\times 10^{-1}$&+1.718$\times 10^{-2}$&
-2.809$\times 10^{-4}$&+1.696$\times 10^{-6}$&-1.720$\times 10^{-2}$&+1.234$\times 10^{-4}$&---\\\hline
$R_V$&+1.491$\times 10^{+0}$&-1.971$\times 10^{-2}$&+5.447$\times 10^{-4}$&
-4.561$\times 10^{-6}$&---&-6.255$\times 10^{-3}$&+9.064$\times 10^{-5}$&-3.187$\times 10^{-7}$\\\hline
$A_V$&+1.933$\times 10^{-1}$&+3.484$\times 10^{-2}$&-9.172$\times 10^{-4}$&
+6.999$\times 10^{-6}$&---&+5.762$\times 10^{-3}$&-6.097$\times 10^{-5}$&+1.929$\times 10^{-7}$\\\hline
$A_{V_i}$&+2.207$\times 10^{-3}$&+5.253$\times 10^{-3}$&-1.970$\times 10^{-4}$&
+2.043$\times 10^{-6}$&---&-5.014$\times 10^{-4}$&+1.898$\times 10^{-6}$&---\\\hline
$R_S$&+8.599$\times 10^{-1}$&-5.657$\times 10^{-3}$&+8.884$\times 10^{-5}$&
+7.253$\times 10^{-7}$&---&+1.024$\times 10^{-2}$&-4.166$\times 10^{-5}$&---\\\hline
$A_S$&+9.477$\times 10^{-1}$&+5.097$\times 10^{-3}$&+1.201$\times 10^{-4}$&
-2.824$\times 10^{-6}$&---&-1.255$\times 10^{-2}$&+4.597$\times 10^{-5}$&---\\\hline
$R_{SO}$&+8.293$\times 10^{-1}$&+3.098$\times 10^{-2}$&-7.747$\times 10^{-4}$&
+6.035$\times 10^{-6}$&---&-3.894$\times 10^{-3}$&+1.799$\times 10^{-5}$&---\\\hline
$R_{SO_i}$&-1.132$\times 10^{-1}$&-5.916$\times 10^{-4}$&+3.596$\times 10^{-6}$&
---&---&+4.458$\times 10^{-3}$&-4.652$\times 10^{-5}$&+1.521$\times 10^{-7}$\\\hline
$A_{SO}$&+9.239$\times 10^{-1}$&+3.091$\times 10^{-2}$&-7.702$\times 10^{-4}$&
+5.982$\times 10^{-6}$&---&-1.874$\times 10^{-2}$&+1.576$\times 10^{-4}$&-4.161$\times 10^{-7}$\\\hline
$R_C$&+3.604$\times 10^{+0}$&-2.103$\times 10^{-1}$&+7.753$\times 10^{-3}$&
-8.155$\times 10^{-5}$&---&+1.074$\times 10^{-1}$&-6.348$\times 10^{-4}$&---\\\hline
$R_{C_i}$&+3.404$\times 10^{-1}$&-1.038$\times 10^{-1}$&+1.294$\times 10^{-3}$&
---&---&+4.501$\times 10^{-2}$&-3.729$\times 10^{-4}$&+9.467$\times 10^{-7}$\\\hline
\end{tabular}
\caption{Model parameters for the subject of this work, the
WP global optical potential
Each term
is a 5 to 7 term separable polynomial in $A$ are $E$ which are
given in Eqs.~\ref{start}-\ref{finish}. Tools have been developed
to facilitate the use of this potential including an on-line optical
potential calculator~\cite{applet}.
}
\label{T1}
\end{table*}

Parameters in this OMP which are at variance with the other OMPs are the
real surface amplitude (Eq.~\ref{eq7}) (which is small), 
imaginary asymmetry ($N-Z$)
(Eqs.~\ref{eq5},\ref{eq9},\ref{eq13}), geometric asymmetry
(Eqs.~\ref{eq16},\ref{eq20},\ref{finish}), and the magic number
dependent terms (Eqs.~\ref{eq3},\ref{eq6}). The asymmetric parameters
and their ramifications are discussed in Sec.~\ref{sec:iso}. The
magic number terms attempts to
better characterize the bonding that occurs in these closed shell nuclei.
Our short term coulomb radius is also untraditionally energy dependent, 
the rational for this will be discussed in Sec.~\ref{sec:iso}.
The twenty three parameters, of Eqs.~\ref{start}-\ref{finish},
which describe the thirteen potential terms of Eq.~\ref{WS},
are listed in Table~{\ref{T1}}.

This potential was put into a standard optical potential
calculator which solves the Schr\"odinger
equation for spin $\frac{1}{2}$-spin $0$ scattering using a 
distorted Born wave approximation (DWBA) in a coulomb wave function basis.  
TALYS~\cite{talys,talys_web} was used which applies ECIS~\cite{ecis} 
to calculate the solution once the final product was developed.
We have produced a Java applet~\cite{applet} which contains our own
optical potential calculator as well as some useful input files
for use in TALYS and ECIS which will let researchers produce results quickly.
The gestation of the parameters will be described in 
Sec.~\ref{sec:opt} but first
an overview will be given of the theoretical structure of 
this and the other recent global OMPs.

\subsection{Theoretical comparison with other global optical potentials}
The general design 
of all three OMPs under study (KD~\cite{globalK}, MD~\cite{globalM}, WP) 
is the same. They 
use similar Woods-Saxon 
functional forms which include volume and spin-orbit terms. Additionally
the KD and WP optical potentials include surface terms.
The KD and MD potentials also include an additional coulomb correction
term, in the WP optical potential the short term coulomb potential
includes energy dependence to include these coulomb correction effects 
(this will be discussed in great detail in Sec.~\ref{subsec:lane}).
Both the WP and MD produce one potential that is utilized for both
proton and neutron scattering while the KD potential has 
separate potentials for the isospin dependence of the projectile.

The KP OMP~\cite{globalK} imaginary parameters are determined by a
dispersive relationship which is dependant on the 
difference between the projectile laboratory energy and the Fermi energy
of the target potential as well as constants designated as part of the real potential.
Of the three optical potentials discussed
it has the deepest theoretical underpinnings and given these
dispersive constraints and high accuracy it has successfully pushed the 
theoretical development of the
global optical model potential to a new level.

A goal of the MD OMP~\cite{globalM}
is to describe the elastic scattering
data sets using fewer parameters.
It uses constants, linear, and the occasional 
quadratic forms to  describe all of its 
Woods-Saxon parameters. It is quite impressive that a high quality isospin 
dependent potential was
produced with so few parameters.

In contrast this work (WP OMP) uses quadratic, cubic, and occasionally 
quartic polynomials
that have no direct relationship to formal scattering theory.
It has the highest amount of adjustable
parameters of the three potentials.
What
it achieves is ease of use, good asymmetry and mirror nuclei analysis,
and complete separability between the
nucleon ($A$) and energy ($E$) parameters. This WP potential also includes
a direct 
imaginary vector isospin asymmetry ($N-Z$) terms which neither of the other  
potentials
have. In Sec.~\ref{sec:iso} a detailed comparison of the isovector differences
of these three optical potentials will ensue.

\section{Procedure}
\label{sec:opt}
This research tried ambitiously to minimize a $\chi^2$ 
(proportional to the
square of the difference between the theoretical fit
and the experimental data).
on over three 
hundred different nucleon-nucleus experiments by adjusting the polynomial fit
to the twenty 
three parameters given in Table~\ref{T1}. To attempt to decouple the terms
from each other, the twenty-three values were varied using a systematic method
detailed below. A listing and discussion of the experimental dataset used to
constrict the variables then follows.
 
\subsection{Calculation Techniques}
\label{subsec:opt}
The fitting elastic nucleon-nucleus scattering code was developed by two 
of the authors (S.~Weppner and R.~Penney).
A numerov routine found in Ref.~\cite{melkanoff} was used to solve the 
non-relativistic position space Schr\"odinger equation in a coulomb basis
with a relativistic correction found in Ref.~\cite{Ca40p80ra}. The 
routine which produced the coulomb wave functions was found in
Ref.~\cite{barnett}. A Powell routine, adopted from Numerical Recipes~\cite{NR},
was used to minimize a weighted 
$\chi^2$.

The code was developed to run on a multi-processor 
parallel system. Each  
processor was assigned at least one nucleon-nucleus 
experimental data set
at a given energy and nucleon number ($A$). This data set
could be as simple
as one experiment or many 
experiments including the observables of total neutron cross section,
total reaction cross section, differential cross section, and polarization. Each
set contained, if available, 
both proton and neutron observables and 
varying target 
proton numbers that all shared a common target
nucleon number and projectile energy.
For example $A=40$, $E=40$ MeV experimental data exists 
for a proton striking $^{40}$Ca and producing
a total reaction cross section observable, a differential cross section, 
and polarization.
These were fit simultaneously with data that exists for a  
neutron striking $^{40}$Ca and producing
a differential cross section, a total reaction cross section, a total 
cross section, and finally also including a proton at 40 MeV striking
$^{40}$Ar and producing a reaction and differential cross section. 
All eight of these experiments made a complete 
working data set in which 
the parameters were varied and the minimization routine for the weighted $\chi^2$
was executed.
By analyzing many different data sets together at the same energy and nucleon
number it was  intended to
reduce systematic error and ultimately derive a better global fit among
chains of target nuclei.

Each fitting cycle was comprised of a three step process.
First all parameters were adjusted except three (the two magic number
and the asymmetry coulomb parameters: Eqs.~\ref{eq3},\ref{eq6},\ref{finish} 
were held fixed).
Since the other twenty parameters
were being adjusted simultaneously it was 
important to analyze only a 
subset of the entire experimental dataset to avoid ambiguities within 
the parameter space. The
datasets used encompassed either 
only $N=Z$ targets or certain sets in which a variety of different $Z$ targets
existed for
a fixed nucleon number and fixed projectile energy
or certain sets in which both neutron and proton 
projectile observables  existed for that
fixed $A$ and fixed projectile energy $E$. 
These sets therefore either had zero $N-Z$ dependence or this dependence
was clearly delineated by including a variety of elements as targets and/or 
projectiles. 
There were 115 data sets of fixed $E$ and $A$ experiments
on which these twenty parameters
were allowed to vary while searching for a minimum weighted $\chi^2$. 
The best values for the parameters were then fit to the 
polynomials of $A$ and $E$ listed in Eqs.~\ref{start}-\ref{finish}.

Second all data sets with non-zero
$(N-Z)$ dependence were used to adjust thirteen parameters.
There were thus ten parameters held fixed which 
were the non-asymmetric potential parameters that had 
other asymmetric terms within the same Woods-Saxon potential term 
(Eqs.~\ref{start},\ref{eq4},\ref{eq8},\ref{eq10},\ref{eq12},\ref{eq15},\ref{eq19},\ref{eq22}
and the magic number terms Eqs.~\ref{eq3},\ref{eq6} were held fixed), 
the other thirteen terms were
varied.
There were 90 experiment data sets 
(each data set might contain more than one experiment as detailed
above) with 
some $N-Z$ dependence that were used for this task. 
The best values for the parameters were then fit to the
polynomials in $A$ and $E$.
The parameters that were free to be modified in both 
of these steps were deduced using an
average of both values which produced 
the minimum weighted $\chi^2$ results.

Lastly, the magic number terms were adjusted (Eqs.~\ref{eq3},\ref{eq6} were 
varied) to find the
minimum weighted $\chi^2$ 
keeping all other parameters
fixed. For the magic numbers the traditional
2,8,20, and 28 were used. There were 40 data sets which contained at least 
one target magic nuclei experiment 
which were used to constrain these last two  parameters.

The variances in the parameters
were minimized to a weighted $\chi^2$  which favored
forward angles over backward angles (in the $\chi$ calculation there was
a square root relationship such that the fiftieth forward angle point
in a differential cross section was weighted only about 
one seventh that of the extreme forward point following Ref.~\cite{globalB}).
Differential cross sections were
favored over polarizations by a factor of 1.5
and in general neutron total cross section point were favored to
be approximately equal to  half of a complete typical differential cross section.
Used, but favored at half the weighting of the total neutron cross section,
were the neutron and proton reaction
cross sections. Choosing the correct weightings was found to be an art
form where a
balance tenuously existed in which every reaction was regarded but certain
reactions were strengthened so that the $\chi^2$ parameter space
contained large relative minima which the search functions could find easily.

To find these minima a Monte Carlo preliminary gross search was done using 
a Sobol~\cite{sobol}
number generator.
Since the parameter space was often twenty dimensions 
this might include up to $3 \times 10^5 $ vectors in which weighted $\chi^2$ were 
first calculated by solving the Schr\"odinger equation for the given potential. 
The lowest ten vectors were analyzed in more fine detail by
seeking the local minima in their vicinity within the $\chi^2$ parameter space.
The $\chi^2$ 
minimization program sought the steepest
derivative in the multi-dimensional parameter space~\cite{NR}.
Although the overall quality of the fits were examined occasionally, the entire
process was close to automatic. 

At first the twenty three parameters had large variances
which were slowly reduced following the three steps described as
the weighted normalized per point
chi-squared was reduced and approached a global minimum gracefully.
The highest 5\% of the $\chi^2$ were thrown out after each fit (which routinely were
the same sets) and then each parameter was fit originally
to a cubic polynomial in $A$ (nucleon number) and $E$ (projectile energy).
Eventually many of the cubics were reduced to quadratics if warranted and
in the asymmetry terms the $A$ polynomial was increased to fourth order to
give a better fit over the entire nucleon target atomic number range.

\subsection{Experimental Datasets}
\begin{table*}
\setlength{\extrarowheight}{2pt}
\begin{tabular}{||c|>{\raggedright\arraybackslash}b{1.85in}|%
|c|>{\raggedright\arraybackslash}b{1.85in}|%
|c|>{\raggedright\arraybackslash}b{1.85in}||}
\hline
\multicolumn{6}{||c||}
{\bf Proton-Nucleus Experimental Elastic and Total Reaction Data References}\\
\hline
\hline
Nucl.&[Ref] (Energies (MeV); Observables) &
Nucl.&[Ref] (Energies (MeV); Observables) &
Nucl.&[Ref] (Energies (MeV); Observables) \\
\hline
$^{12}$Be&\cite{Be12p55a}(55;dcs)&
$^{12}$C&\cite{C12p35po}(20-84;dcs,pol)\ \cite{Fe54p30rca}(30;dcs,pol) 
&$^{13}$C&\cite{Fe54p30rca}(30;dcs,pol)\ \cite{C12p35a3}(35;dcs)
\\ \cline{1-2}
$^{14}$N&\cite{Ar40p30a}(30;dcs)\ \cite{C12p35a3}(35;dcs)&&
\cite{C12p40ra}(30-60;rcs) \cite{C12p35a3}(35;dcs)
&&\cite{C13p35po}(35;dcs,pol)
\\
&\cite{N14p50a,Ar40p30a}(50;dcs,pol) \ \cite{Ar40p40ra}(35;rcs)&&
\cite{C12p35a2}(35;dcs)\ \cite{C12p40a}(40;dcs)
&& \cite{C13p72po,C13p72po2}(72;dcs,pol)
\\
&\cite{N14p122a}(122;dcs)\ \cite{N14p142a}(142;dcs)&&
\cite{C12p49a}(49;dcs) 
\ \cite{C12p50a}(50;dcs) 
&&\cite{C13p135po}(135;dcs,pol)
\\ \cline{1-2} \cline{5-6}
$^{15}$N& \cite{C12p35a3}(35;dcs)\ \cite{N15p44rca}(39,44;dcs)&&
\cite{C12p61a}(61;dcs) 
\ \cite{C12p65a}(65;dcs,pol) 
&$^{16}$O&\cite{Fe54p30rca}(30;dcs,pol)\ \cite{O16p35a}(35;dcs,pol)
\\ & \cite{O18p42rc}(42,44;dcs) \ \cite{C13p72po2}(65;dcs,pol)&&
\cite{C12p65ra}(65;rcs) 
\cite{Ca40p152po}(75,150;dcs,pol)&&
\cite{C12p35a3,N15p44rca}(35;dcs) \ \cite{O16p46ra}(30-47;rcs)
\\ \cline{1-2} 
$^{17}$O&\cite{C12p35a3}(35;dcs)\ \cite{O16p66a}(66;dcs)&&
\cite{C12p142ra}(81-180;rcs) \ \cite{C12p96a}(96,dcs) &&  
\cite{O16p43a}(43,46;dcs) 
\\ \cline{1-2} 
$^{18}$O& \cite{C12p35a3}(35;dcs)\ \cite{O18p42rc}(42,44;dcs)&&
\cite{C12p122a}(122,160;dcs,pol)&&
\cite{C12p50a,N14p50a}(49;dcs,pol)
\\
&\cite{O18p43ra}(43;dcs)\ \cite{O16p66a}(67;dcs)&& 
\cite{C12p135po}(135;dcs,pol,rcs)&&
\cite{Ne20p65a}(65;dcs,pol)\ \cite{C12p65ra}(65;rcs) 
\\ \cline{1-2} 
$^{20}$O& \cite{O20p30a}(30;dcs)\ \cite{O18p43ra}(43;dcs)&& 
\cite{C12p142a}(142;dcs)  
 \cite{C12p145poa}(145;dcs,pol)&&
\cite{O16p135po}(135;dcs,pol)
\\ \cline{1-2} \cline{5-6}
$^{20}$Ne& \cite{C12p35a3}(35;dcs)\ \cite{Ar40p40ra}(35;rcs)&&
\cite{C13p72po2}(150;dcs pol)&
$^{22}$O&\cite{O22p47ra}(47;dcs)
\\ \cline{5-6}
& \cite{Ne20p65a}(65;dcs,pol)&&
\cite{C12p156ra}(156;dcs,pol)&
$^{22}$Ne& \cite{C12p35a3,N15p44rca}(35;dcs)
\\\cline{1-2}\cline{3-4}\cline{5-6}
$^{24}$Mg&\cite{Mg24p45a}(30-45;dcs)\ \cite{Mg24p50a}(50;dcs,pol)&
$^{28}$Si$^*$&\cite{Si28p52rc}(52;dcs)&
$^{29}$Si&\cite{C13p72po}(72;dcs,pol)
\\\cline{3-6}
&\cite{C12p65a,Ne20p65a}(65;dcs,pol)&
$^{30}$Si&\cite{Si30p52a}(52;dcs)&
$^{31}$P&\cite{C13p72po}(72;dcs,pol)
\\ \cline{3-4}\cline{5-6}
&\cite{Mg24p80a}(80;dcs,pol)\ \cite{Mg24p135a}(135;dcs,pol) &
$^{32}$Si&\cite{Si32p42a}(42;dcs)&
$^{32}$S&\cite{S32p53a}(53;dcs)\ \cite{C12p65a}(65;dcs,pol) 
\\ \cline{1-6}
$^{34}$S&\cite{S34p30a}(30;dcs)&
$^{34}$Ar&\cite{S32p53a}(47;dcs)&
$^{36}$Ar&\cite{Ar42p33a}(33;dcs)
\\ \cline{1-6}
$^{37}$Cl&\cite{N15p44rca,C12p35a3}(35;dcs)&
$^{39}$K&\cite{N15p44rca,C12p35a3}(35;dcs)&
$^{40}$Ar&\cite{Ar40p30rca}(30,33,37,41;dcs,pol)
\\ \cline{1-4} 
$^{40}$Ca$^*$&\cite{C12p142ra}(80-180;rcs) \ \cite{Ca40p45a}(30-48;dcs)&
$^{42}$Ar&\cite{Ar42p33a}(33;dcs)&&
\cite{Ar40p30a,N15p44rca,C12p35a3}(30-50;dcs,pol)
\\ \cline{1-4}
$^{42}$Ca&\cite{Ca40p45a}(30-48;dcs)\ \cite{Ca42p40ra}(30-48;rcs)&
$^{44}$Ar&\cite{Ar42p33a}(33;dcs)&
&\cite{Ar40p40ra}(36-47;rcs) \ \cite{Ne20p65a}(65;dcs,pol) 
\\ \cline{3-6} 
&\cite{C12p35a3}(35;dcs)\ \cite{Fe54p50ra}(49;dcs)&
$^{44}$Ca&\cite{Ca40p45a}(30-48;dcs)\ \cite{Ca42p40ra}(30-48;rcs)&
$^{45}$Sc&\cite{N15p44rca}(35;dcs)\ \cite{Fe54p50ra}(50;dcs,pol)
\\ \cline{5-6}
&\cite{Ca42p65po}(65;dcs,pol)&
&\cite{Fe54p50ra}(49;dcs) \ \cite{Fe54p65ra,Ne20p65a}(65;dcs,pol) &
$^{46}$Ti&\cite{Ne20p65a,Ca42p65po}(65;dcs,pol)\ \cite{Ti46p100rc}(100;dcs)
\\ \cline{1-6}
$^{48}$Ca&\cite{Ca40p45a}(30-48;dcs)\ \cite{Ca42p40ra}(30-48;rcs)&
$^{48}$Ti&\cite{Ti48pra}(30-48;rcs)&
$^{50}$Ti&\cite{Ca42p65po}(65;dcs,pol)\ \cite{Ti46p100rc}(100;dcs)
\\ \cline{5-6} 
&\cite{Fe54p65ra,Ne20p65a,Ca42p65po}(65;dcs,pol)&
&\cite{Ne20p65a,Ca42p65po}(65;dcs,pol)\ \cite{Ti46p100rc}(100;dcs)&
$^{50}$Cr&\cite{Si28p52rc}(52;dcs)\ \cite{Ca42p65po}
\\ \cline{1-2} \cline{5-6}
$^{52}$Cr&\cite{Ca42p65po}(65;dcs;pol)&
&\cite{Al27prc}(100;rcs)&
$^{54}$Cr&\cite{C12p35a3}(35;dcs)\ \cite{Ca42p65po}(65;dcs;pol)
\\ \cline{1-2} \cline{5-6}
$^{54}$Fe$^*$&\cite{Fe54p30a}(30,40,62;dcs)&
&\cite{C12p156ra}(156;dcs)&
$^{56}$Fe$^*$&\cite{C12p40ra}(40,60;rcs)\ \cite{Fe56p65a2}(65;dcs,pol)
\\ \cline{3-6}
&\cite{Fe54pra}(30-48;rcs)\ \cite{Fe54p40ra}(40;dcs,pol)&
$^{58}$Fe&\cite{C12p35a3}(35;dcs) &
$^{59}$Co&\cite{Fe56p30rca}(30;dcs)\ \cite{Co59p40rc}(40;dcs) 
\\
&&
&&
&\cite{Fe54pra}(30,40;rcs)\ \cite{Ni58p65ra}(65;dcs,pol)
\\ \hline
\end{tabular}
\caption{This is a partial listing of the proton-nucleus experimental
data used in the fitting process of this work, the remainder is located
in Table 7 of Ref.~\cite {globalK}. The references are listed followed
by the laboratory energy of the projectile and the observables found
in the reference article (rcs - total reaction cross section, 
dcs-differential cross section, pol-analyzing power). The data
has an energy range from 30 MeV to 160 MeV,  and a nucleon number,
$12\le A \le 60$ as required by the calculation. 
Table 7 of Ref.~\cite{globalK} contains$^*$
a substantial listing of experimental data references primarily for the
$^{27}$Al, $^{28}$Si, $^{40}$Ca, $^{54}$Fe, $^{56}$Fe, $^{58}$Ni
and $^{60}$Ni, all those experimental references were 
also used in this work if they were 
within this optical potentials applicable energy range.   }
\label{T2}
\end{table*}

\begin{table*}
\setlength{\extrarowheight}{2pt}
\begin{tabular}{||c|>{\raggedright\arraybackslash}b{1.85in}|%
|c|>{\raggedright\arraybackslash}b{1.85in}|%
|c|>{\raggedright\arraybackslash}b{1.85in}||}
\hline
\multicolumn{6}{||c||}
{\bf Neutron-Nucleus Experimental Elastic and Total Reaction Data References}\\
\hline
\hline
Nucl.&[Ref] (Energies (MeV); Observables) &
Nucl.&[Ref] (Energies (MeV); Observables) &
Nucl.&[Ref] (Energies; (MeV) Observables) \\
\hline
$^{12}$C&\cite{C12n40ra}(30-49;rcs) \ \cite{C12n35a}(35;dcs)  &
$^{14}$N&\cite{C12n40ra}(30-49;rcs)&
$^{16}$O&\cite{C12n40ra}(30-49;rcs)\ \cite{O16p135po}(35;dcs)
\\ \cline{3-6}
& \cite{Fe56n55a}(55-75;dcs)\ \cite{C12n156a}(65-156;dcs)&
$^{27}$Al$^*$&\cite{C12n40ra}(30-49;rcs)&
$^{28}$Si$^*$&\cite{deb_misc2}(misc.;rcs)
\\ \cline{3-6}
&\cite{C12n96a}(96;dcs) \ \cite{deb_misc,deb_misc2}(misc.;rcs)&
$^{40}$Ca$^*$&\cite{C12n40ra}(30-49;rcs)\ \cite{C12n156a}(65-156;dcs)&
$^{56}$Fe$^*$&\cite{deb_misc2}(misc.;rcs)
\\ \hline
\end{tabular}
\caption{This is a partial listing of the neutron-nucleus experimental
data used in the fitting process of this work, the remainder is located
in Table 1 of Ref.~\cite {globalK}. The references are listed followed
by the laboratory energy of the projectile and the observables found
in the reference article (rcs - total reaction cross section,
dcs-differential cross section). The data
has an energy range from 30 MeV to 160 MeV,  and a nucleon number,
$12\le A \le 60$ as required by the calculation. 
Table 1 of Ref.~\cite{globalK} contains$^*$
a substantial listing of experimental data references primarily for
$^{24}$Mg,$^{27}$Al, $^{28}$Si, $^{32}$S,
$^{40}$Ca, and $^{56}$Fe,
all those experimental references were also used in this work 
if they were
within this optical potentials applicable
energy range. The neutron total cross section data
sets of Refs.~\cite{finlay1,finlay2} were also extensively used.   }
\label{T3}
\end{table*}

This work strived for a comprehensive collection of  experimental
data sets. As a starting point the excellent summary of elastic
data found in Ref.~{\cite{globalK}} was used, the product of the KD optical
potential group. This was
supplemented with new data and also nuclei outside the assumed range
of the KD
potential. These additional data sets  were either nuclei lighter than $^{27}$Al or
those nuclei which are non-spherical and were not considered in that article, there
were also a few additional data sets that were discovered and added to our
database.
All proton-nucleus data sets used that were not in Ref.~{\cite{globalK}}
are listed in
Table~\ref{T2}. The smaller set of neutron-nucleus data sets used
here and not cited in  Ref.~{\cite{globalK}} are listed in Table~\ref{T3}.
A reader of this work, Ref.~{\cite{globalK}}, and the
growing EXFOR/CSISRS database at
the {\it National Nuclear Data Center}~\cite{tin_data} have a near
complete compilation of elastic nucleon-nucleus and total cross section
experimental data listings at
intermediate energies.

Not all the experimental 
data was used to help constrain this OMP. 
Some data, usually published
before 1960, had large systematic differences with later reactions and were
disregarded. Also, as mentioned in Sec.~\ref{sec:opt}, calculations
were only done if angular differential data existed at a given energy
and nucleon number. It was established for this fitting procedure
that total cross section data did
not have enough information (only one data point) 
to constrain the parameters adequately
so it was only used in conjunction with
differential experimental data. Occasionally if the energy of
a total cross section was close to an energy where differential 
data existed, the total reaction cross section data was adjusted following
the forms given in Refs.~\cite{Deb,Deb2}. This procedure
was not needed for the total neutron cross section data of 
Refs.~\cite{finlay1,finlay2} since the energy coverage was substantial
for these datasets.

\section{Results} 
\label{sec:results}
To produce the calculations shown in this section the DWBA scattering code
TALYS~\cite{talys,talys_web} was used for consistency. All experimental
data is shown as black circles, the new results of this
work are depicted as  light solid green lines (WP OMP), the calculations of 
Koning and Delaroche~\cite{globalK} (KD OMP)
are shown using dark blue dashed lines and the calculations of 
Madland ~\cite{globalM} (MD OMP)
are depicted using a medium red dot-dashed lines. The motivation
is not to be exhaustive but to give a fair representative overview
of the features of these three modern optical potentials. As a global
summary it can be  concluded that that all three potentials do a 
fairly good job fitting the presently available
elastic scattering data and dramatic
contrasts
are not proffered until Sec.~\ref{sec:iso}. The details of this section
do illustrate
some minor differences and so first their will be an examination of
the neutron-nucleus observables then following with the proton-nucleus 
observables.
\begin{figure*}
\includegraphics*[width=7.0in,angle=0]{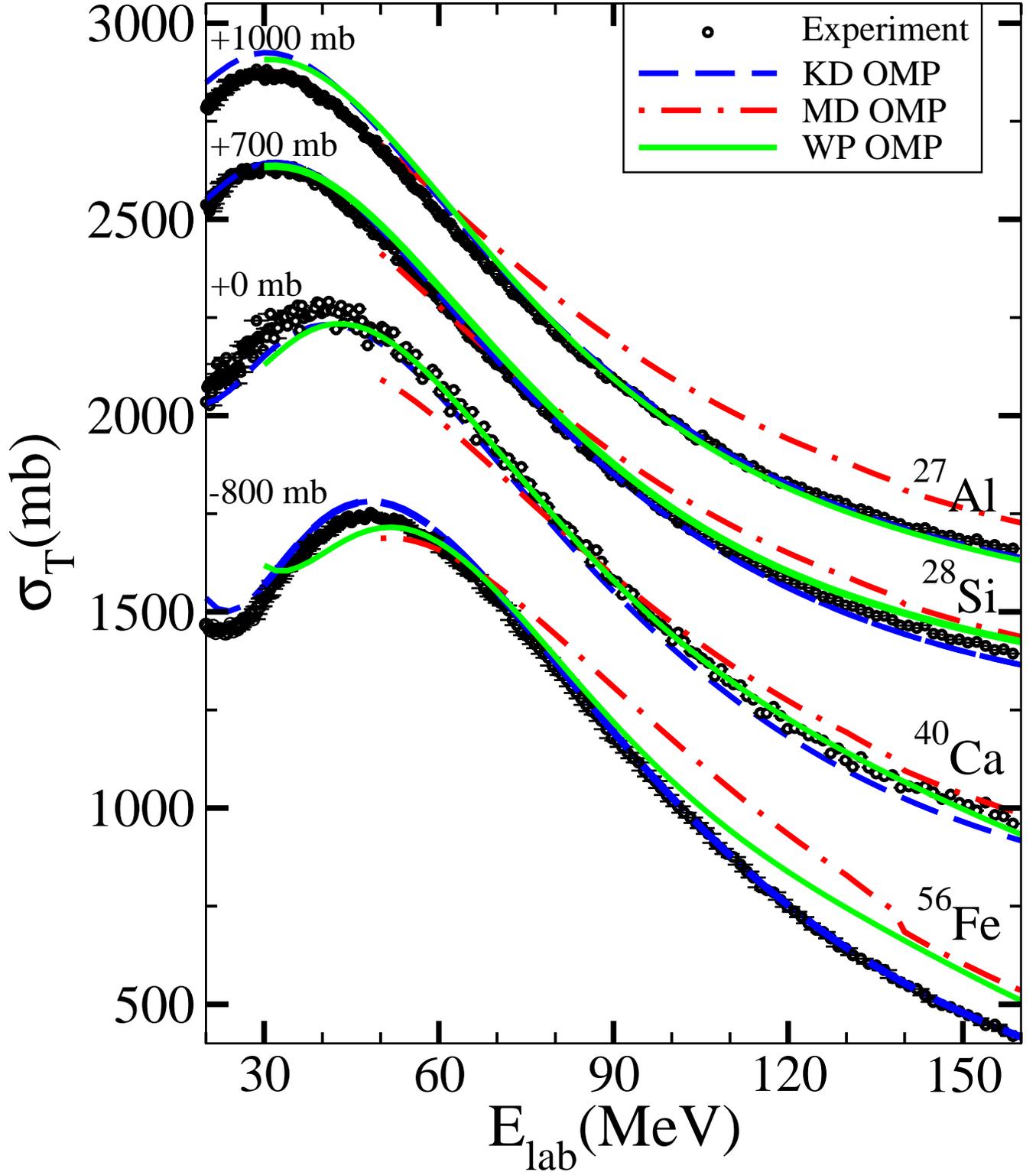}
\caption{(Color online) The experimental total cross section for neutrons 
scattering from 
$^{27}$Al\cite{finlay1},$^{28}$Si\cite{finlay1},$^{40}$Ca\cite{finlay2} and 
$^{56}$Fe\cite{finlay2} from 20 MeV to 160 MeV for the laboratory
energy of the neutron. They are
fit to three different optical potential calculations
(KD\cite{globalK},MD\cite{globalM} and to this work: WP) which
are described in the legend. The experimental data and theory have
been offset by constant amounts for multiple comparisons
on one graph. The WP
and MD calculations have a minimum energy limit of 30 and 50 MeV 
respectively.}
\label{ncs1}
\end{figure*}
\begin{figure*}
\includegraphics*[width=7.0in,angle=0]{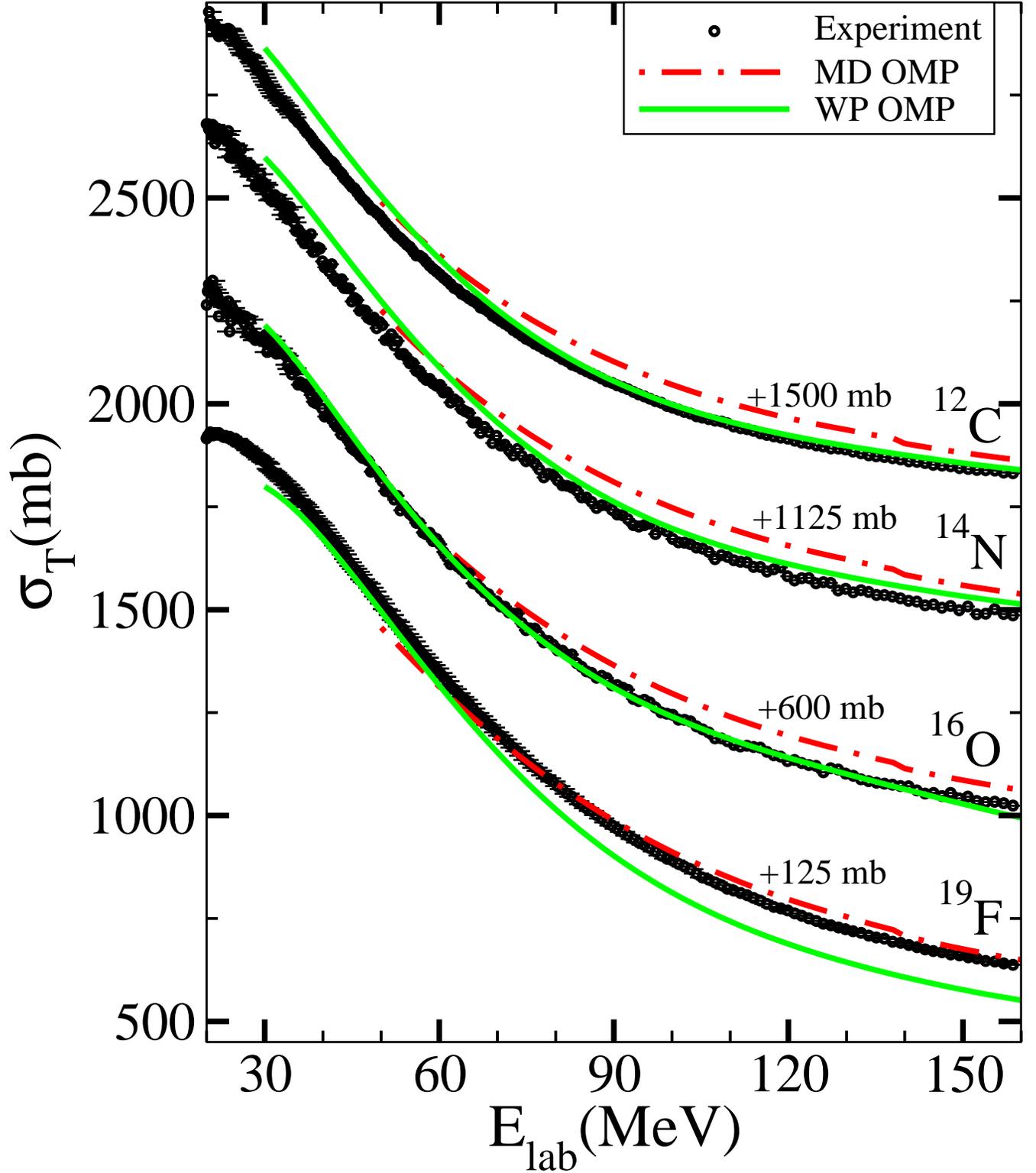}
\caption{(Color online) The experimental total cross section for neutrons scattering from
the light nuclei of 
$^{12}$C\cite{finlay2},$^{14}$N\cite{finlay1},$^{16}$O\cite{finlay1} and
$^{19}$F\cite{finlay2} from 20 MeV to 160 MeV for the laboratory
energy of the neutron. They are
fit to two different optical potential calculations
(MD\cite{globalM} and to this work: WP) which
are described in the legend. 
The experimental data and theory have
been offset by constant amounts for multiple comparisons
on one graph. The WP
and MD calculations have a minimum energy limit of 30 and 50 MeV
respectively. The KD calculation is not applicable for these light nuclei.}
\label{ncs2}
\end{figure*}
\begin{figure*}
\includegraphics*[width=7.0in,angle=0]{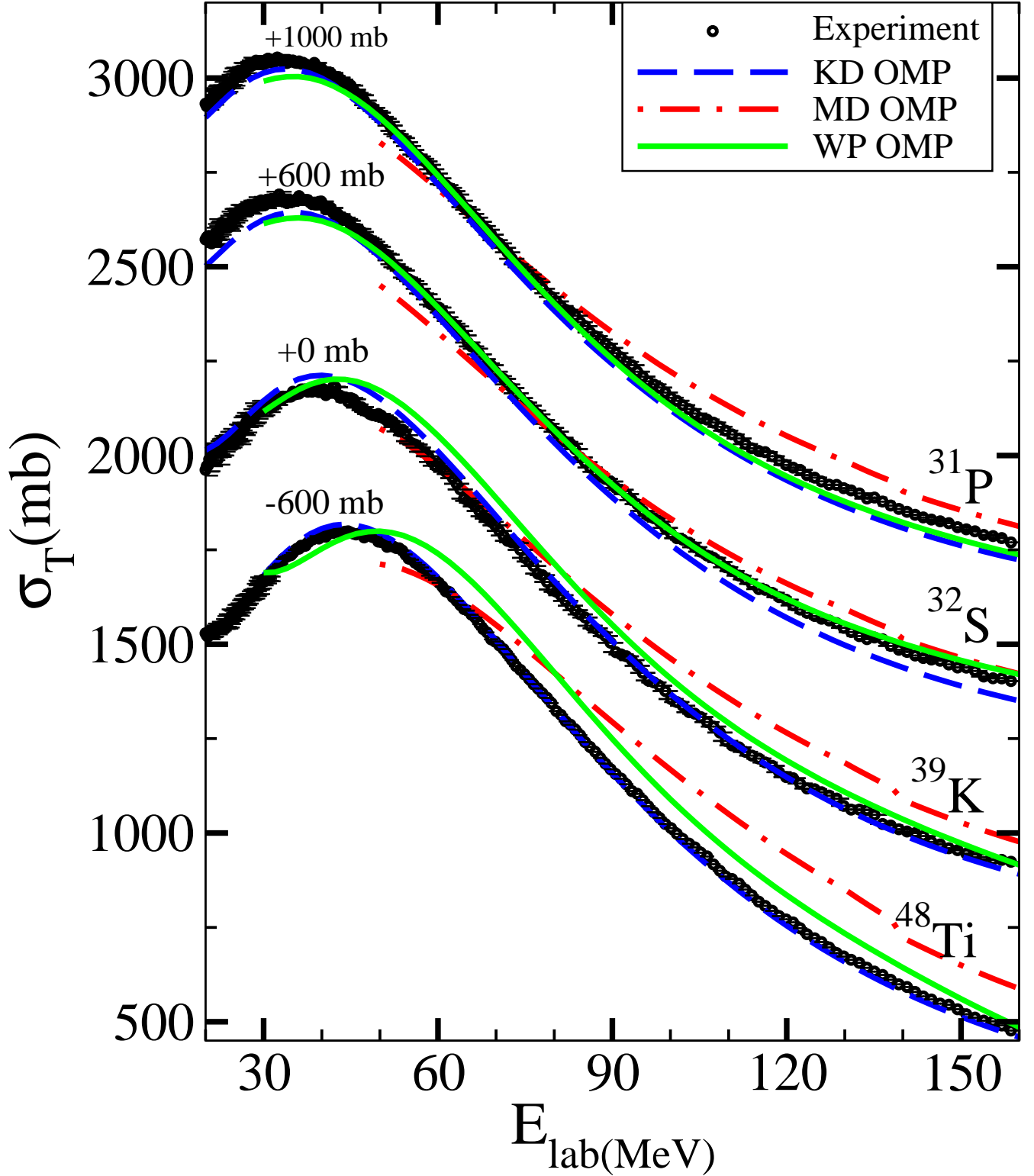}
\caption{(Color online) The experimental total cross section for neutrons scattering from
$^{31}$P\cite{finlay2},$^{32}$S\cite{finlay2},$^{39}$K\cite{finlay2} and
$^{48}$Ti\cite{finlay2} from 20 MeV to 160 MeV for the laboratory
energy of the neutron. They are
fit to three different optical potential calculations
(KD\cite{globalK},MD\cite{globalM} and to this work: WP) which
are described in the legend. The experimental data and theory have
been offset by constant amounts for multiple comparisons
on one graph. The WP
and MD calculations have a minimum energy limit of 30 and 50 MeV
respectively.}
\label{ncs3}
\end{figure*}
\begin{figure*}
\includegraphics*[width=7.0in,angle=0]{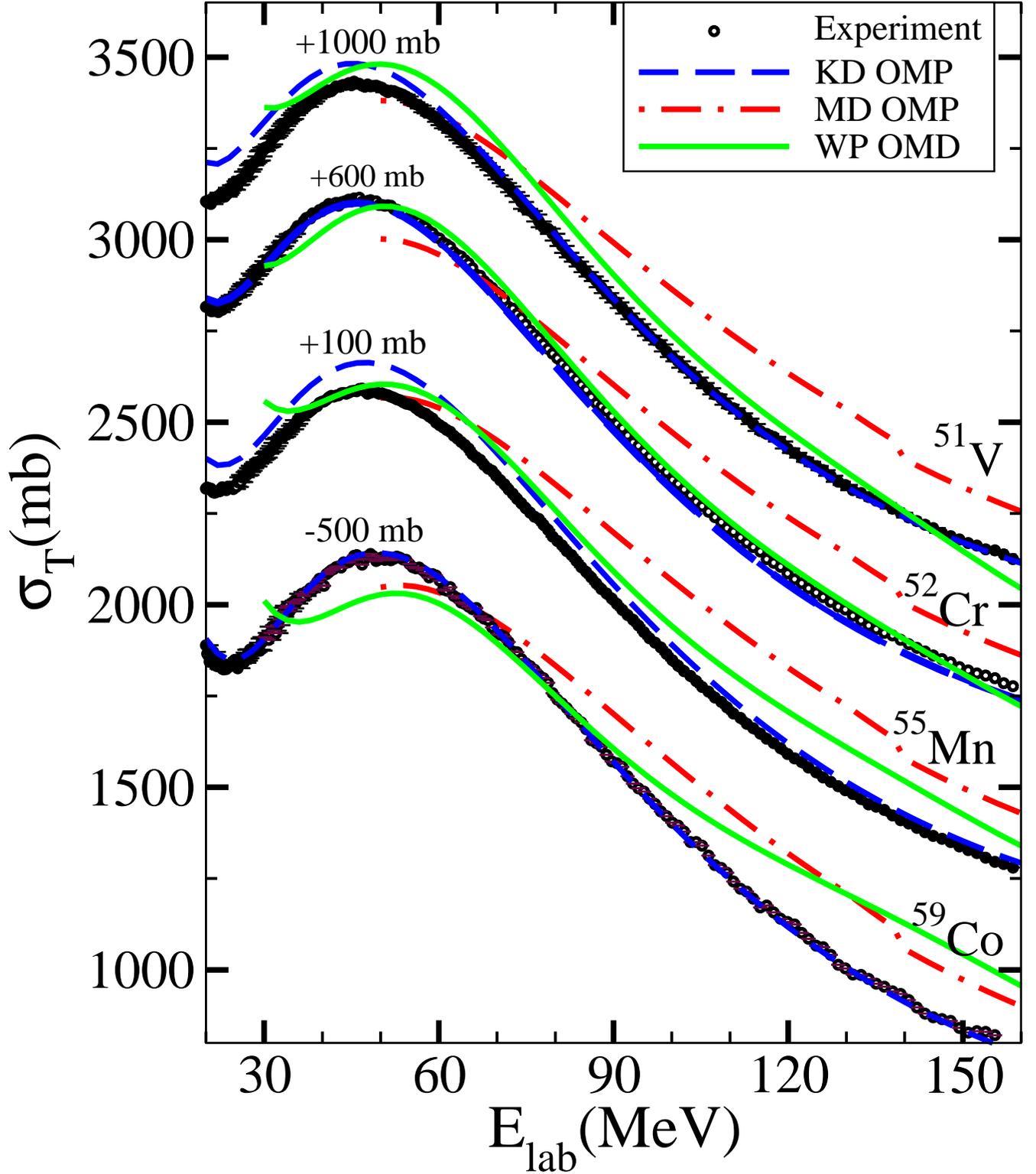}
\caption{(Color online) The experimental total cross section for neutrons scattering from
$^{51}$V\cite{finlay2},$^{52}$Cr\cite{finlay2},$^{55}$Mn\cite{finlay2} and
$^{59}$Co\cite{finlay2} from 20 MeV to 160 MeV for the laboratory
energy of the neutron. They are
fit to three different optical potential calculations
(KD\cite{globalK},MD\cite{globalM} and to this work: WP) which
are described in the legend. The experimental data and theory have
been offset by constant amounts for multiple comparisons
on one graph. The WP
and MD calculations have a minimum energy limit of 30 and 50 MeV
respectively.}
\label{ncs4}
\end{figure*}

\subsection{Neutron-Nucleus Observables}
In Figs.~1-4 the calculations of
total neutron cross section results to the comprehensive data sets
of Refs.~\cite{finlay1,finlay2} are compared. Overall the calculations of 
the KD OMP of Ref.~\cite{globalK} do the best job at
reproducing the experimental data, usually within a remarkable
1\%. The potential of this work (WP OMP) is usually within 5\% of the 
experimental data sets and that of the MD OMP is usually within 10\%. These are
plotted on an elongated linear scale to accentuate the disparity but it
should be recognized that even a 10\% difference between theory and experiment 
is extraordinary
for most nuclear scattering observables.

The remarkable fit by the KD OMP calculation is justified, the authors
considered this data important and thus weighted
it accordingly. They also had, as discussed in Sec.~{\ref{sec:theory}},
a separate proton nucleus and neutron nucleus potential. At 
energies greater than 50 MeV, where neutron-nucleus elastic scattering data 
is scarce, this was one of the few observables used to fit 
their neutron optical potential.
With fewer constraints
this observable, with its high caliber of data~\cite{finlay1,finlay2}, was easier
to fit. Conversely the optical potential of this work
and the optical potential of the MD OMP had to continually
compromise by both fitting 
neutron-nucleus and proton-nucleus observables simultaneously.

In Fig.~\ref{ncs1} the calculations are fit to four standard
nuclei. As in all these
calculations the WP calculation of this work (solid green line) was not
tested below 30 MeV and the MD calculation was not used below 
50 MeV. Likewise the KD calculation is not applicable to the lighter 
nuclei ($A<27$) targets. 
Systematic trends emerge where the KD calculation
runs lower and closer to the experimental results
than the other two calculations for the
neutron-nucleus total cross section. 
The WP calculation is 
also systematically lower and closer than the MD 
calculation for the neutron total
cross section observables
as typified in this figure. There is a small kink in
the MD results at E=130 MeV, this is a real discontinuity in the
Woods-Saxon function form parameters for this potential
at both 130 MeV and 140 MeV. The history
of the phenemological fitting endeavor shows that researchers have struggled
to fit these higher energies well, the present works are no exception. 

Fig.~\ref{ncs2} contains the lighter targets which follow the same trends
as described for  Fig.~\ref{ncs1}. Of the two applicable optical potential 
calculations (the KD OMP is defined only when $A\ge 24$) the WP OMP
of this work does better in all cases except for
neutron scattering from $^{19}$F. The odd nuclei have non-zero spin forces
which are not included in the solution technique which in part may explain 
the anomaly.
\begin{figure*}
\includegraphics*[width=7.0in,angle=0]{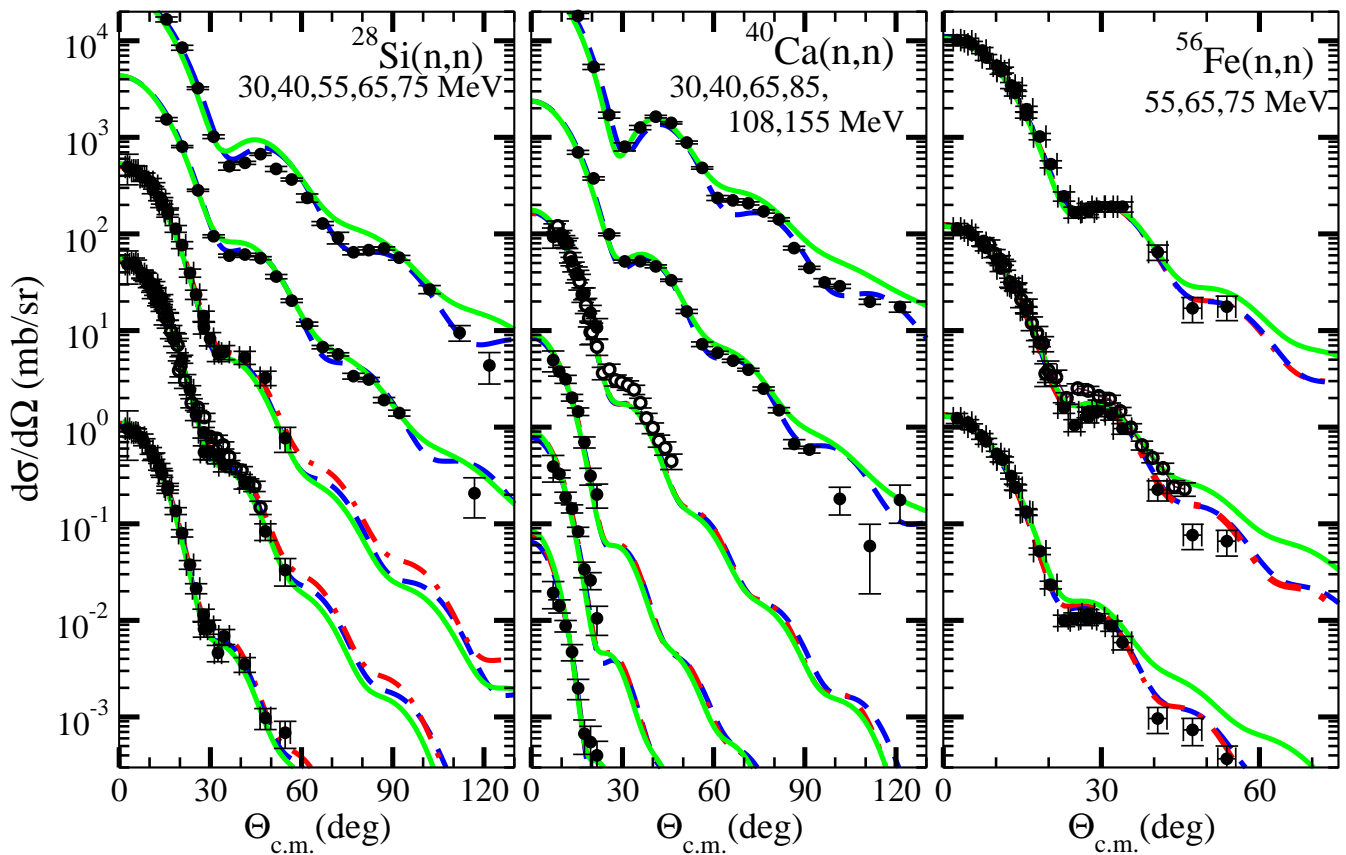}
\caption{(Color online) Neutron-nucleus elastic differential cross section experimental
data for the target nuclei of
$^{28}$Si\cite{Si28n40a,Fe56n55a,Fe56n65a},
$^{40}$Ca\cite{Ca40n40a,Fe56n65a,C12n156a},
$^{56}$Fe\cite{Si28n40a,Fe56n55a,Fe56n65a,Fe56n65a2} at a variety of incident
laboratory energies. They are
fit to three different optical potential calculations
(KD\cite{globalK},MD\cite{globalM} and to this work: WP).
As in all figures, KD OMP is a blue dashed line, MD OMP is a red dot-dashed
line and WP OMP is a green solid line.
The experimental data and theory have
been offset by constant amounts for multiple comparisons
on one graph. 
The target labels at the top read left to right
correspond to the calculations and experimental data read top
to bottom. 
The
MD calculations has a minimum energy limit of 50 MeV
and is therefore missing from calculations below that energy.}
\label{ndcs}
\end{figure*}

Concluding the results of neutron-nucleus total cross section calculations
are Fig.~\ref{ncs3},\ref{ncs4} which contain a variety of less common
nuclei and odd spin nuclei neutron total cross sections. The KD optical
potential of Ref.~\cite{globalK} which was explicitly not fit to these  
non-spherical nuclei still does
remarkably well, the same trends discussed in the earlier figures
still hold. What becomes apparent in the heavier nuclei ($50\le A\le 60$)
calculations of this work (WP OMP - solid light green line) is that there seems
to be a systematic  energy shift; the shape of the curve is good, but it seems to 
contain a small shift towards higher energies, a possible explanation will 
be proffered in Sec.~\ref{subsec:lane}.  

The neutron-nucleus differential experimental data is scarce above
30 MeV (see Table~\ref{T3} and Ref.~\cite{globalK}). 
A representative sample of the data with 
calculations is shown in Fig.~\ref{ndcs}. Overall all three potentials
describe the experimental data adequately, most impressive
are the results of the calculation of the MD potential which simultaneously
fits both proton-nucleus and neutron-nucleus data
while also
using the fewest terms and parameters of the three optical potentials examined.
Systematically the WP potential of this work has the weakest results, 
especially when the scattering angle is greater than 45 degrees 
This implies
that to improve these results the WP potential would need to give more weighting
to higher angle results than was determined (see Sec.~\ref{sec:opt} which
discusses the weightings chosen).
To adequately measure the effects and form of the isospin 
dependent and asymmetry terms in the optical
potential which is used by the WP and MD OMPs more high energy
neutron-nucleus differential and reaction data is sorely 
needed~\cite{dispersive1,neutron2}.

Overall the best optical potential calculation for neutron-nucleus
scattering is the work of Koning and Delaroche (KD OMP)~\cite{globalK}. It has
the best fit to the total cross section observables (even the non-spherical and
odd nuclei which were not used to constrain that potential) and it also
does an admirable job with the differential observables.
It has a wide energy range and is thus suitable for
systematic neutron-nucleus  studies. Its largest deficit is that it
is not applicable to light nuclei ($A<24$) which are important in 
astrophysics and biological physics. 

\subsection{Proton-Nucleus Scattering Observables}
The proton-nucleus elastic scattering observables will now be considered. The
potential used by the MD and WP calculations remains the same while the
KD potential uses a different optical potential to calculate the proton-nucleus
observables.
\begin{figure*}
\includegraphics*[width=7.0in,angle=0]{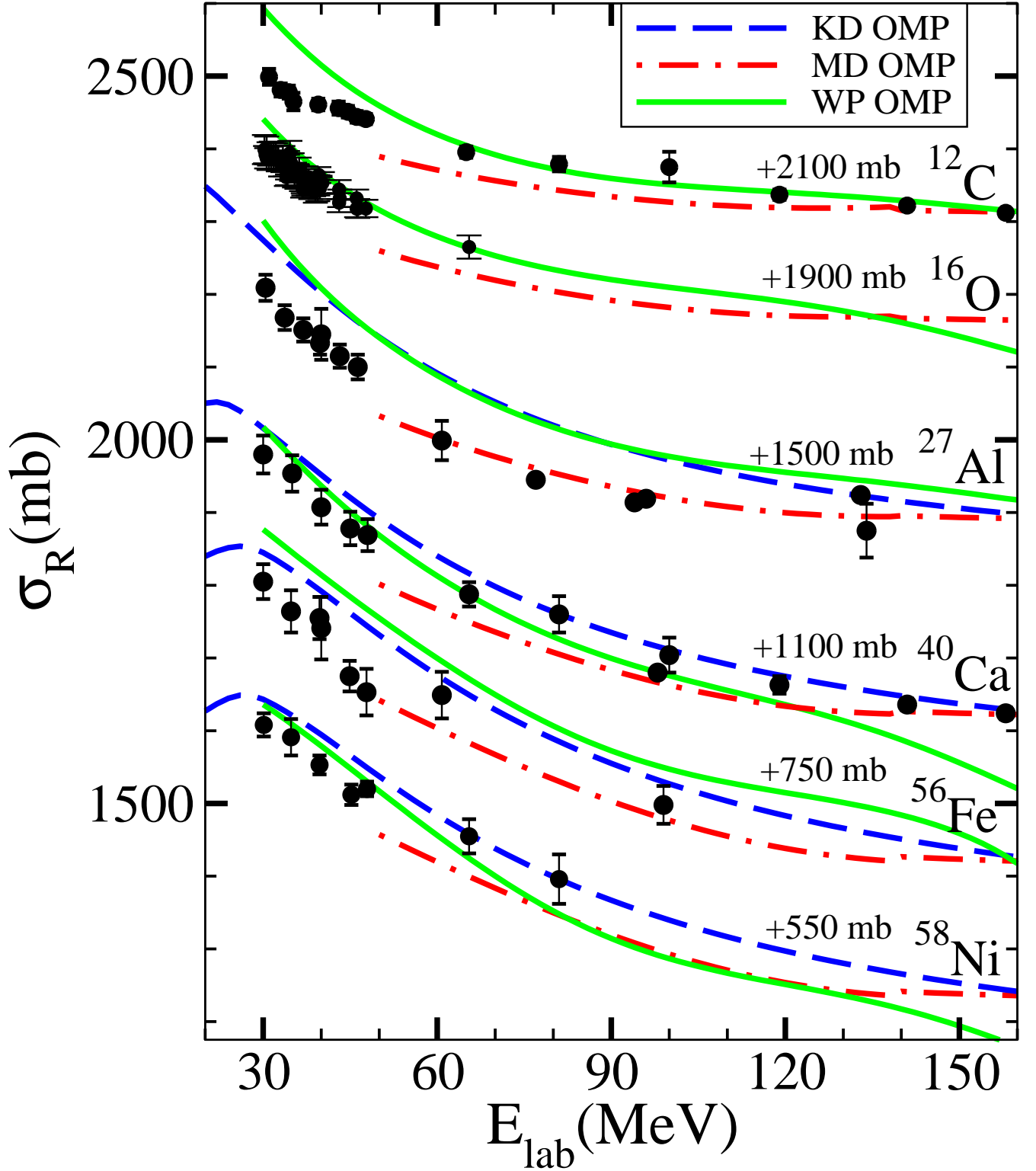}
\caption{(Color online) Proton-nucleus total inelastic cross section data for the targets
$^{12}$C,$^{16}$O,
$^{27}$Al,$^{40}$Ca,
$^{56}$Fe, and $^{58}$Ni
 from 20 MeV to 160 MeV for the laboratory
energy of the projectile proton. The experimental data comes from a variety of
sources and are compiled and discussed
in Refs.~{\cite{Al27prc,Deb,globalK}}. There are also
new higher energy
measurements for $^{12}$C and $^{40}$Ca found in Ref.~{\cite{C12p142ra}}.
They are
fit to three different optical potential calculations
(KD\cite{globalK},MD\cite{globalM} and to this work: WP) which
are described in the legend. The experimental data and theory have
been offset by constant amounts for multiple comparisons
on one graph. The WP
and MD calculations have a minimum energy limit of 30 and 50 MeV
respectively.}
\label{rpcs}
\end{figure*}
\begin{figure*}
\includegraphics*[width=7.0in,angle=0]{diffcs.protons.new.eps}
\caption{(Color online) Proton-nucleus elastic Rutherford reduced
differential cross section experimental
data for the target nuclei of
$^{27}$Al\cite{N15p44rca,C12p61a,Al27p62a,Al27p142ra,C12p156ra},
$^{28}$Si\cite{C12p65ra,Si28p100a},
$^{40}$Ca\cite{Ca40p45a,C12p61a,Ca40p80ra},
at varying proton
laboratory energies. Refer to Fig.~\ref{ndcs} for details of the legend for the 
theoretical 
calculations.}
\label{pdcs}
\end{figure*}
\begin{figure*}
\includegraphics*[width=7.0in,angle=0]{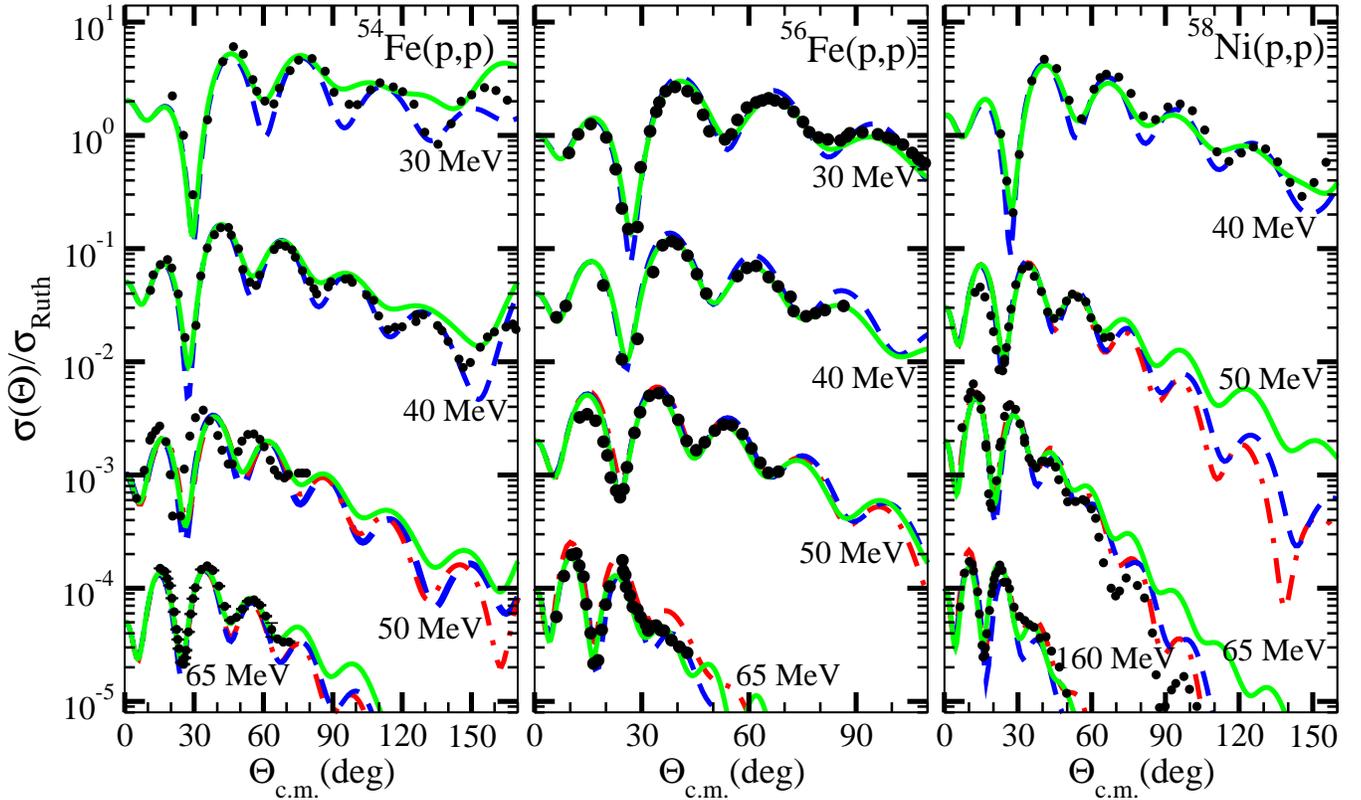}
\caption{(Color online) Proton-nucleus elastic Rutherford reduced
differential cross section experimental
data for the target nuclei of
$^{54}$Fe\cite{Fe54p30rca,Fe54p40ra,Fe54p50ra,Fe54p65ra},
$^{56}$Fe\cite{Fe56p30rca,Fe56p40ra,Fe54p50ra,Ni58p65ra},
$^{58}$Ni\cite{Ni58p40ra,Ni58p65ra,Ni58p100ra,Ni58p160ra},
at varying proton 
laboratory energies. Refer to Fig.~\ref{ndcs} for details of the legend for the 
theoretical
calculations.}
\label{pdcs2}
\end{figure*}
\begin{figure*}
\includegraphics*[width=7.0in,angle=0]{diffcs.protons3.new.eps}
\caption{(Color online) Proton-nucleus elastic Rutherford reduced
differential cross section experimental
data for the target nuclei of
$^{34}$Ar\cite{S32p53a,S30p53ra},
$^{30}$S\cite{S30p53ra},
$^{40}$Ar\cite{Fe54p65ra},
$^{45}$Sr\cite{Fe54p50ra},
$^{40,42,44,48}$Ca\cite{Fe54p65ra,Ca42p40ra},
$^{50,52,54}$Cr\cite{Ca42p40ra},
at varying proton
laboratory energies. 
The target labels at the top read left to right
correspond to the calculations and experimental data read top
to bottom for the two rightmost panels.
Refer to Fig.~\ref{ndcs} for details of the legend for the 
theoretical
calculations.}
\label{pdcs3}
\end{figure*}
\begin{figure*}
\includegraphics*[width=7.0in,angle=0]{diffcs.protons4.new.eps}
\caption{(Color online) Proton-nucleus elastic Rutherford reduced
differential cross section experimental
data for the target nuclei of
$^{12,13}$C\cite{C12p135po,C13p135po},
$^{16}$O\cite{O16p135po,O16p43a},
$^{18,20,22}$O\cite{O18p43ra,S30p53ra,O22p47ra},
$^{28}$Si\cite{Si28p100a},
$^{40,42}$Ca\cite{Fe56p30rca,Ca40p45a},
$^{40,42,44}$Ar\cite{Ar40p30rca,Ar42p33a},
at varying proton
laboratory energies. 
The target labels at the top read left to right
correspond to the calculations and experimental data read top
to bottom for the two leftmost panels.
Refer to Fig.~\ref{ndcs} for details of the legend for the theoretical
calculations.}
\label{pdcs4}
\end{figure*}

A representative sample of the inelastic or reaction total cross section
data are shown in Fig.~\ref{rpcs} with projectile
energies ranging from 30 MeV to 160 MeV. The calculations in this case are not
as close to the experimental data, and the data
itself is much sparser and more unsure than the earlier
neutron total cross section data. More accurate high energy data is needed in
this observable to better constrain the optical model theory (it
plays a significant role in determining the
absorptive strength\cite{C12p142ra,reaction2}),
an example will be proffered to illustrate
this in Sec.~\ref{sec:iso}.
All three
global optical potentials do well but are not excellent.
As with
the neutron total cross sections the legend is the same and
the figure is elongated and plotted
on a linear scale to emphasis the differences.

\begin{figure*}
\includegraphics*[width=7.0in,angle=0]{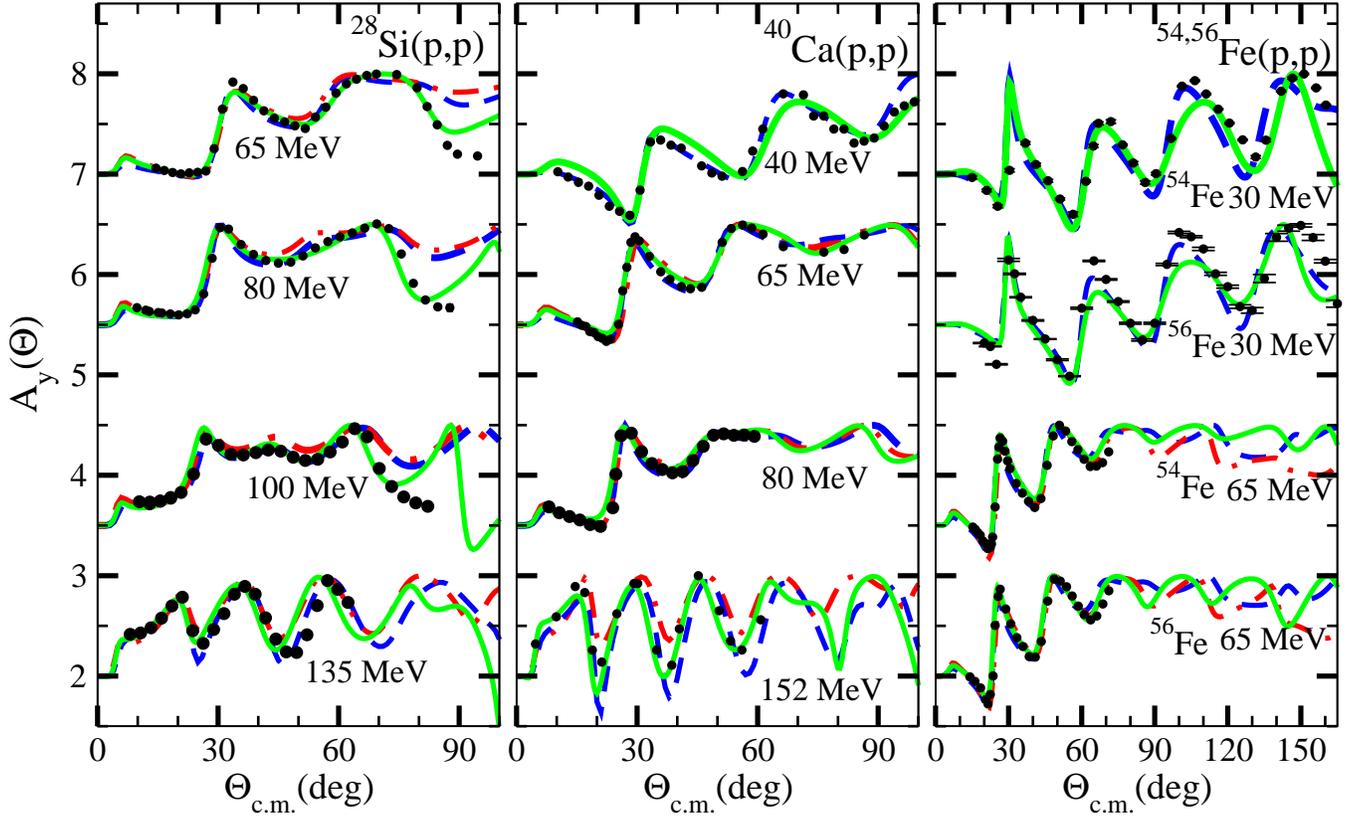}
\caption{(Color online) Proton-nucleus elastic 
analyzing power spin observable experimental
data for the target nuclei of
$^{28}$Si\cite{Fe54p65ra,Si28p100a},
$^{40}$Ca\cite{Ca40p40po,Fe54p65ra,Ca40p80po,Ca40p152po},
$^{54,56}$Fe\cite{Fe54p30rca,Fe56p30po,Fe54p65ra,Ni58p65ra},
at varying proton
laboratory energies. Refer to Fig.~\ref{ndcs} for details of the theoretical
calculations.}
\label{ppo1}
\end{figure*}

\begin{figure*}
\includegraphics*[width=7.0in,angle=0]{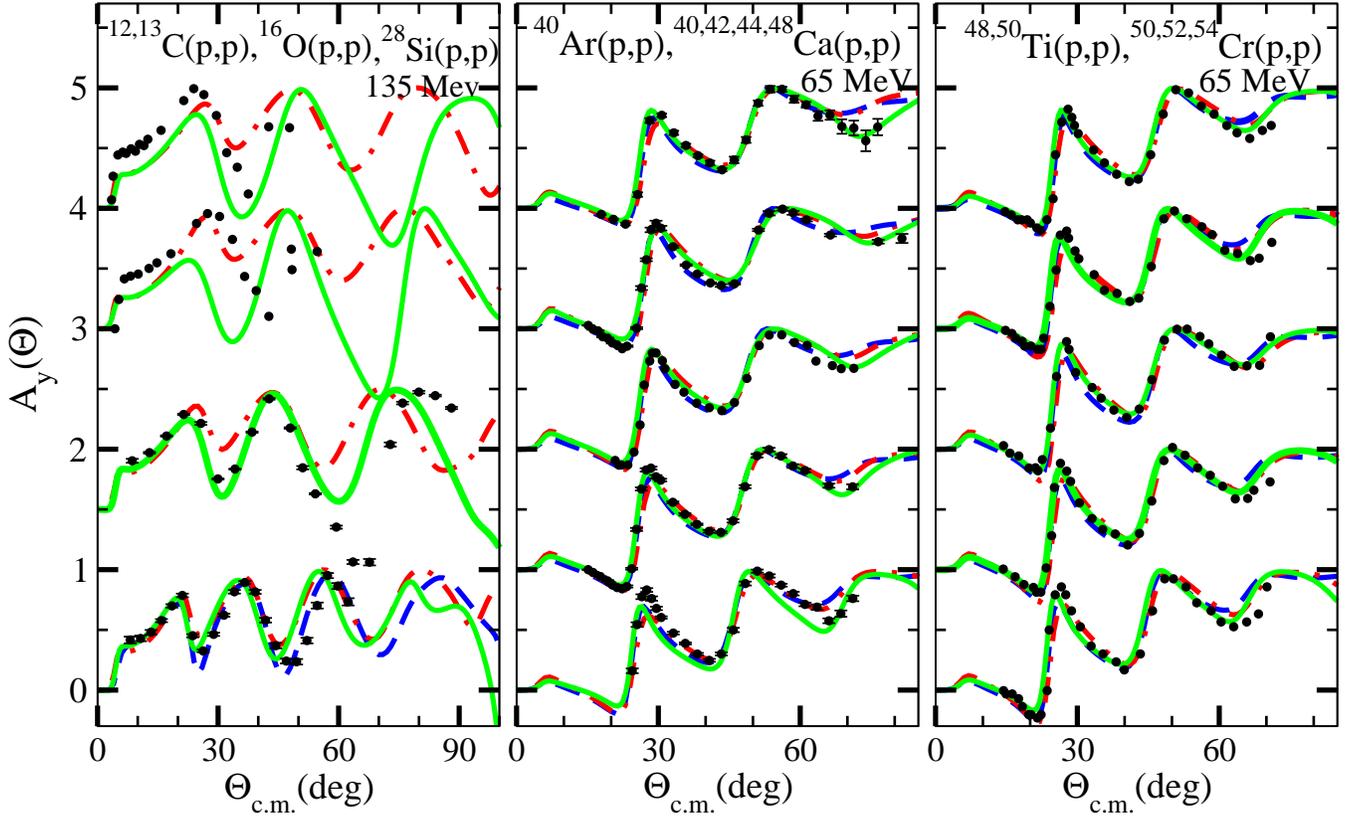}
\caption{(Color online) Proton-nucleus elastic
analyzing power spin observable experimental
data for the target nuclei of
$^{12,13}$C\cite{C12p135po,C13p135po},
$^{16}$O\cite{O16p135po},
$^{28}$Si\cite{Si28p100a},
$^{40,42,44,48}$Ca\cite{Fe54p65ra,Ca42p65po},
$^{40}$Ar\cite{Ne20p65a},
$^{48,50}$Ti\cite{Ne20p65a},
$^{50,52,54}$Cr\cite{Ca42p65po},
at varying proton
laboratory energies. 
The target labels at the top read left to right
correspond to the calculations and experimental data read top
to bottom. 
Refer to Fig.~\ref{ndcs} for details of the theoretical
calculations.}
\label{ppo2}
\end{figure*}

\begin{figure*}
\includegraphics*[width=7.0in,angle=0]{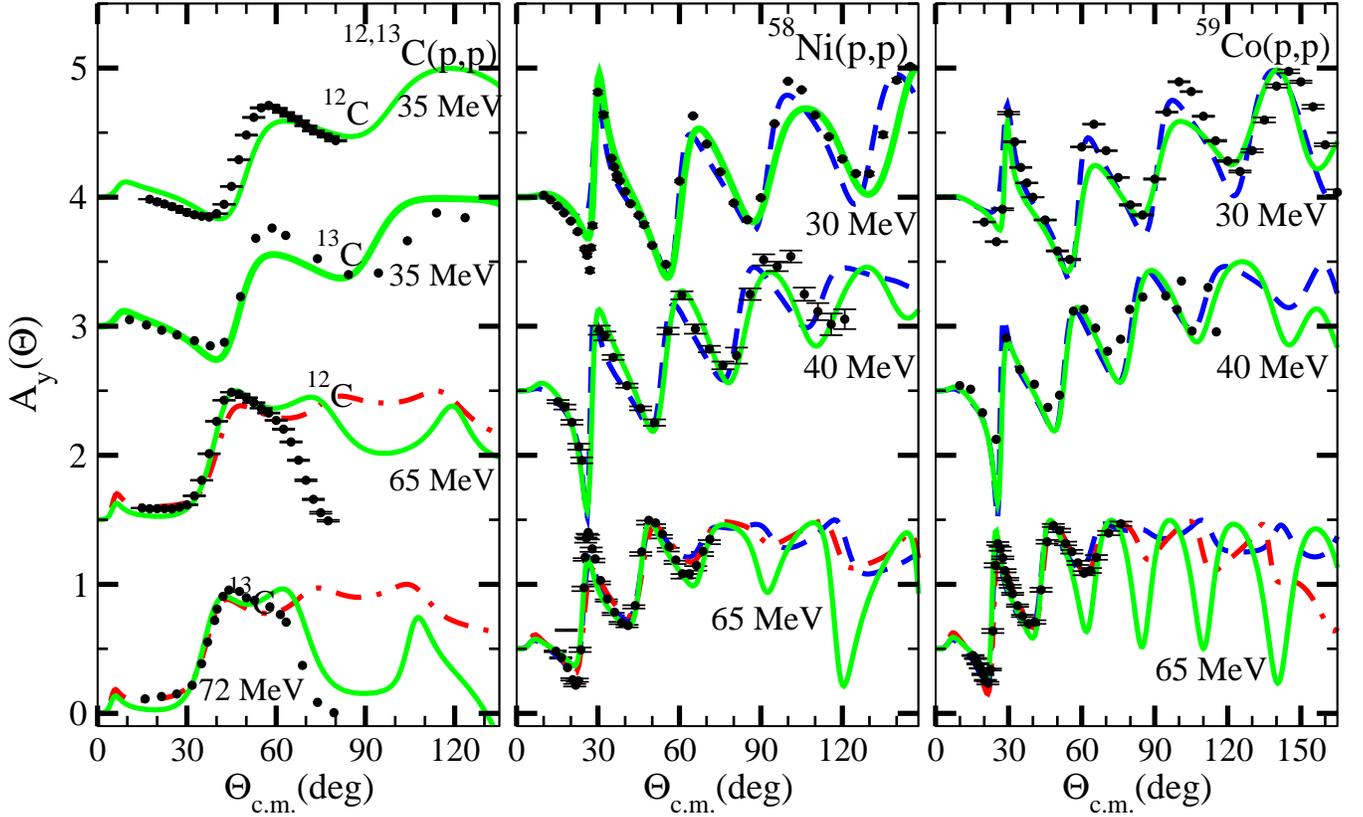}
\caption{(Color online) Proton-nucleus elastic
analyzing power spin observable experimental
data for the target nuclei of
$^{12,13}$C\cite{C12p35po,C13p35po,C12p35po,C13p72po},
$^{58}$Ni\cite{Ni58p30po,Ni58p40ra,Ni58p65ra},
$^{59}$Co\cite{Fe56p30po,Co59p40rc,Ni58p65ra},
at varying proton
laboratory energies. Refer to Fig.~\ref{ndcs} for details of the theoretical
calculations.}
\label{ppo3}
\end{figure*}

There is a much more substantial amount of experimental data for the
proton-nucleus differential cross section and 
the spin observable analyzing power and a representative
sample is contained in Figs.~\ref{pdcs}-\ref{ppo3}. 
In Fig.~\ref{pdcs} there is
a comparison of some
of the common nuclei targets, $^{27}$Al, $^{28}$Si, and $^{40}$Ca for
the differential cross section normalized to the coulomb Rutherford
differential cross section.
Then an examination of the heavier common target
subset ($^{54,56}$Fe and $^{58}$N) transpires in Fig.~\ref{pdcs2}.
All these nuclei were calculated in Ref.~\cite{globalK} by the KD calculation
and in this work are compared with the MD and WP potentials (in the 
applicable energy ranges, an MD calculation is not created for 
an  energy of the projectile of  less than 50 MeV). Overall all the
three calculations do well. Some systematic trends become apparent:
the WP OMP struggles at the higher angles to reproduce the 
experimental data (as it did with the neutron-nucleus observables). 
The disappointing fit at larger angles insinuates that the balance of the 
weighting functions for the WP OMP favored the forward
angles too much. 
The KD OMP often exaggerates the minimum which
probably signifies that the coulomb strength is slightly weak short range as 
prescribed in that 
potential (in general
the long term coulomb force mildly obscures the diffraction effect generated
by the short term forces). Because these are plotted on a logarithmic
scale the percent difference between the theory and experiment is often
larger than it appears (sometimes as much as 50\% compared to
10\% for the total cross sections). 
The quality of fit, as ascertained
by an analysis of the $\chi^2$ in 
Ref.~\cite{globalK}, is not as good as with the neutron observables, yet
it is still quite impressive for all three optical potentials.

The next two figures (Fig.~\ref{pdcs3} and Fig.~\ref{pdcs4})
examine the reduced differential cross section of
target nuclei which are non-spherical and
farther from the line of stability than fit in Ref.~\cite{globalK} using
the KD potential. Many of these examples 
are odd non-spin zero nuclei. Overall the three potentials do 
surprisingly well with the same systematic problem as with the more popular 
and common
nuclear targets, however some 
aspects are dissatisfying. In Fig.~\ref{pdcs3} 
the results of
the proton-nucleus reduced differential cross section of the
calcium and chromium isotopes are shown and none of the 
calculations do extremely well describing
all four isotopes simultaneously. The research of this work 
focused on trying to find the strength
of the isovector antisymmetry term by simultaneously fitting these isotope's
observables together. The resulting calculation (the solid green line) fit
$^{40}$Ca and $^{44}$Ca isotopes comfortably but struggle with the
the two other isotopes, the other calculations are in similar predicaments. The 
general shape is correct but the minima and maxima are often missed by over
25\%.
It seems that a asymmetry term which is linear and has mirror 
symmetry evades accurate 
discovery. The same difficulties with describing the
calcium isotopes were recently discussed in Ref.~\cite{dispersive1}. 
These same arguments can be made, to a lesser extent, with chromium as shown
in Fig.~\ref{pdcs3}. The
failure to create an excellent isovector asymmetry term is the motivation for
the analysis of Sec.~\ref{sec:iso}. It also must be recognized that 
these potentials are often also
missing a general spin-spin term which may be important for
non-spin zero targets. As this potential was being developed a spin-spin term
was used but the results had a substantial amount of noise and it was eventually 
removed.
A  competent analysis of the validity and 
strength of this spin-spin term would be a worthy endeavor for future work.

Fig~\ref{pdcs4} contains the reduced differential cross section
of some lighter targets (if $A <{27}$ the KD calculation
is not applicable). This figure shows the struggles the calculations have to fit
at higher energies which
has been a common problem with an optical
model calculation.~\cite{C12p122a}. Likewise the back angles continually
cause difficulty in the lighter targets. Recently there has been success
in fitting these back angles by adding terms which add non-local
approaches to these local OMPs~\cite{cooper,o20} by adding angular momentum 
and parity dependencies to the potential.

In the center panel
are highlighted some newer differential cross section data of
the oxygen isotopes. The WP calculation (which is the
only one applicable) does well in describing the general trend
of the first diffraction minimum shifting as the neutron number
increases. The mirror nuclei of calcium and argon are the
targets in the right most panel. A recurring disappointment in not being
able to fit all the isotopes with the same excellent quality of fit, an important 
motivation
for this work. However, for those working with exotic beams, this 
potential does give a good starting basis for continuing research.


In Figs.~{\ref{ppo1}-\ref{ppo3}} are plotted a representative sample of the
proton-nucleus spin analyzing power (polarization)
observables using the same legend as in all earlier figures. 
In Fig.~\ref{ppo1} which examines
tradition nuclear targets, all the calculations do well. The fitting
of the polarization variable simultaneously with the differential 
cross section involved an interesting tension between the relative
weight of  
the $\chi^2$ of both observables since the polarization
is normalized by the differential cross section. The results show
that in many instances this work (the WP OMP) fit the polarization better than 
it did with the 
differential cross section. In all three polarization figures the 
major difficulty was again with the lighter nuclei, specifically 
the carbon isotopes~\cite{C12p122a,c12deb} as depicted 
in Figs.\ref{ppo2}-\ref{ppo3}. 

The global optical potential of this work is the best in
reproducing this spin observable over a wide range of targets and chains of nuclei.
Additional polarization experimental data is always welcome because it measures
the interference between the central and spin-orbit terms and therefore is a 
good constraint on their relative strengths. The present dataset has a satisfactory
amount of these reactions
at projectile energies lower than 100 MeV but additional higher energy polarization
experiments would be appreciated.

\subsection{Isoscalar Strength}\label{subsec:isoscale}
\begin{figure}[ht]
\includegraphics*[width=3.35in,angle=0]{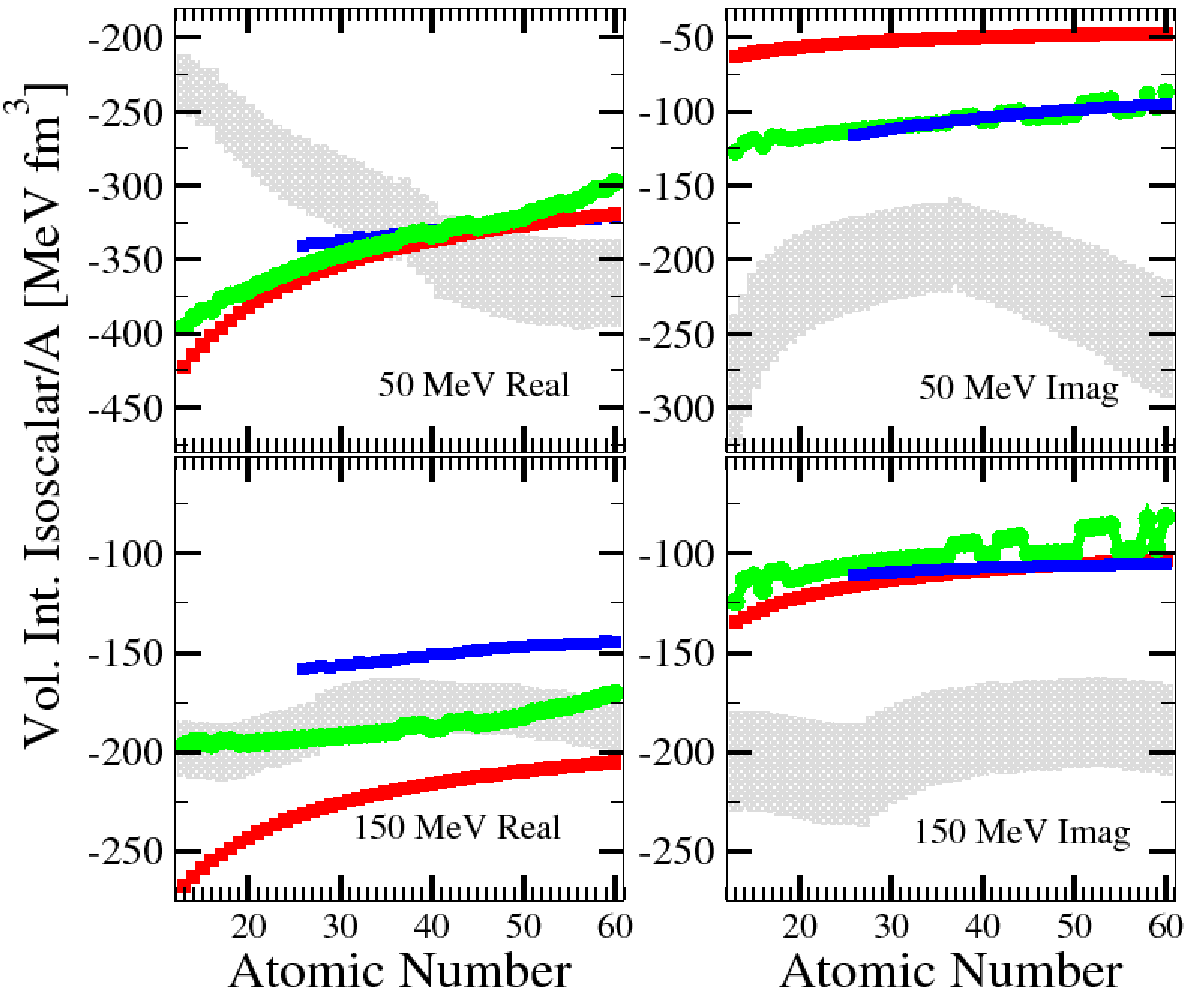}
\caption{(Color online) A figure of theoretical comparisons. Plotted are the
isoscalar volume real and imaginary
comparisons between the three potentials that are under examination at 50 MeV
and 150 MeV projectile energy. The light green, medium blue, and
dark red lines represent the WP, KD, and MD OMPs respectively. The light
gray shaded area represents
typical volumes for microscopic $t-\rho$ potentials 
(as detailed in the text). Each atomic
number has a sample nuclei attache to it. These nuclei are:
$^{13}$C, $^{14}$C, $^{15}$N, $^{16}$N, $^{17}$O, $^{18}$O, $^{19}$F, $^{20}$F,
$^{21}$Ne, $^{22}$Ne, $^{23}$Na, $^{24}$Na, $^{25}$Mg, $^{26}$Mg, $^{27}$Al, $^{28}$Al,
$^{29}$Si, $^{30}$Si, $^{31}$P, $^{32}$P, $^{33}$S, $^{34}$S, $^{35}$Cl, $^{36}$Cl,
$^{37}$Cl, $^{38}$Ar, $^{39}$K, $^{40}$Ar, $^{41}$K, $^{42}$Ca, $^{43}$Ca, $^{44}$Ca,
$^{45}$Sc, $^{46}$Ti, $^{47}$Ti, $^{48}$Ti, $^{49}$Ti, $^{50}$V, $^{51}$V, $^{52}$Cr,
$^{53}$Mn, $^{54}$Fe, $^{55}$Mn, $^{56}$Fe, $^{57}$Co, $^{58}$Ni, $^{59}$Co, $^{60}$Ni.
These same nuclei will be used in many of the remaining figures.}
\label{isoscale}
\end{figure}

To analyze optical potentials it has often been instructive
to calculate the volume integrals of the various terms 
to further illicit theoretical comparison.
In Fig.~\ref{isoscale}  
the central real and imaginary isoscalar volume 
components at 50 MeV and 150 MeV projectile energy are reproduced. 
There is some ambiguity in determining
the isoscalar component for the OMPs discussed here
so the technique used was the 
average between the proton-nucleus and the
neutron-nucleus central volume integrals.
Using this definition the central potential of all traditional OMPs
can be split into the isoscalar and
isovector parts respectively
\begin{equation}
{\cal V}_V(E)= V_0(E)+{\cal I}(E)\label{isoscale0},
\end{equation}
where ${\cal I}(E)$, the isovector component, has its sign dependent on the isospin
of the projectile and $V_0(E)$ is the isoscalar term. 
For Fig.~\ref{isoscale} a representative sample of nuclei 
were chosen that are on or
near the line of stability,
the explicit list of nuclei used in the calculations
are listed in the figure caption.
The green, red, and blue lines are the
WP, KD, and MD OMPs volume integrals of the isoscalar term ($V_0(E)$
respectively and they have 
broad agreement which is encouraging. All three 
potentials, fit independently to experimental data, 
developed roughly the same strength
for the isoscalar central component. 
The only large discrepancy is the MD OMP at low energy
has a much smaller imaginary 
component because it is missing a surface term. The other two 
OMPs at this energy (KD and WP) have a 
significant imaginary surface term which adds to the strength of the
attractive central potential. 
Incidentally if the WP OMP of this work  is examined closely in Fig.~\ref{isoscale} 
protrusions are seen, especially in the 
imaginary panels. These are due to the 
new magic number
term that the potential has included (Eqs.~\ref{eq3},\ref{eq6}).
The ramifications of this 
inclusion are not readily apparent in comparisons
with the reaction data, future studies on 
the effect of closed shells on optical
potential behavior would be fruitful.

To elicit further insight
the volume integrals created from microscopic Watson $t-\rho$ (
nucleon-nucleon scattering matrix folded with target density)
potentials~\cite{watson1,watson2,amos1} are 
plotted in light gray in Fig.~\ref{isoscale}. The
specific apparatus was developed, in part, by earlier work of one of the 
present authors (Weppner) and detailed 
in Refs.~\cite{med1,med2,tmatrix} In brief, the light gray 
shaded regions use a variety of 
nucleon-nucleon potentials~\cite{cdbonn2000,nijmegen,av18} folded with
theoretical densities~\cite{DH2,chiral1} using a variety of techniques. 
These microscopic 
optical potentials have been satisfactory in reproducing experiment 
at the lower energies and better at
the higher energies. Rarely do the microscopic potentials, 
which have different fundamental motivations, match the
quality of these three phenemological potentials in reproducing these observables. 
Significant 
systematic differences are recognized in the isoscalar volume integrals 
especially in the light nuclei and the imaginary
term but there are also regions of general agreement.

In summary, these three OMPs describe the known 
elastic scattering data well and have similar
isoscalar magnitudes. Detailing their advantages;
the KD OMP is well suited for neutron projectile projects,
the MD OMP, in its simplicity, does well 
with differential cross sections, and the WP OMP
has strengths in the exotic nuclei and in the spin observables. 
In the next section
larger theoretical differences will be proffered, while 
examining the isovector components, 
that will eventually lead to quite
disparate scattering observable results.

\section{Isovector Term Analysis}
\label{sec:iso}
The asymmetry isovector term, which measures target neutron and proton imbalance,
is a point of illustrative comparison and will be the focus
of this section. This research attempted to develop an  isospin
consistent potential solely by fitting to elastic experimental reaction data,
here the success of this effort will be assessed.
First the historical ansatz of a simple linear $N-Z$
dependence for this potential term
will be examined, then a Lane analysis check
of the isospin character of this OMP will be
proffered. Finally a detailed comparison of the isovector volume integral of the
three OMPs alongside microscopic
optical potential results
will occur with connections to experiment established.

\subsection{Validity of $N-Z$}\label{subsec:valitity}
Recently the simple linear $N-Z$ 
dependence of the asymmetric potential has been called to question~\cite{globalMR}. 
All three phenemological potentials,
following tradition, have used this standard so it is informative to 
use microscopic
techniques to develop the optical potential from the  nucleon-nucleon
potential to gain insight into the origin of this $N-Z$ term (following 
Refs.~\cite{lane1,satchler2,satchler1}). The 
potential between two nucleons usually contains isospin vector components 
(usually a spin dependent and a spin independent piece, for simplification 
here they are combined). Assuming the valid impulse
approximation at high energies, the techniques of 
Kerman, McManus, Thaler~\cite{KMT} and Watson~\cite{watson1,watson2,med1,med2} 
can be used
to construct
the nucleon-nucleus optical potential from a sum of nucleon-nucleon potentials:
\begin{equation}
V_{asym} \approx\sum_i^A V_{isoNN}(\tau_{proj}\cdot\tau_i),
\end{equation} 
switching over to raising and lowering operators
\begin{eqnarray}
V_{asym} \approx\nonumber \\
\sum_i^A V_{isoNN}
\frac{1}{2}(\tau_{proj_+}\tau_{i_-}+\tau_{proj_-}\tau_{i_+}+2\tau_{proj_z}\tau_{i_z}).
\end{eqnarray}
If the difference of the asymmetry piece
between the proton 
projectile and neutron projectile on the same target is 
calculated then microscopically
it has this format:
\begin{eqnarray}
V_{asym}(t_{proj}=+\frac{1}{2})-V_{asym}(t_{proj}=-\frac{1}{2}) \nonumber \\
\approx V_{isoNN}\big(\frac{N}{2}\tau_{proj_-}-\frac{Z}{2}\tau_{proj_+}\big)\nonumber \\
-V_{isoNN}\big(\frac{N-Z}{4}
(\tau_{proj_z}(+\frac{1}{2})+\tau_{proj_z}(-\frac{1}{2}))\big).
\end{eqnarray}
The first two terms are the simplest 
inelastic charge-exchange terms, often referred to as 
quasi-elastic charge-exchange~\cite{satchler2} because they do not involve
the direct exchange of nucleons and they are between isobaric analog states.
The first inelastic term is exclusively non-zero 
for the proton projectile, the second 
inelastic term
is non-zero exclusively for the neutron projectile
and the last two terms
are the elastic contributions (the same contribution 
occurs for both the proton or neutron projectile). 
A direct 
$N-Z$ factor is derived directly from the elastic scattering component, 
the inelastic term does not contain the equivalent proportionality
(the proton potential contains a non-zero $N$ and the 
neutron potential contains a non-zero
negative $Z$ however
these potentials are not implemented simultaneously). This 
is the first affirmation that
a linear $N-Z$ term in the optical potential, 
which by definition is both refractive (elastic)
and absorptive (inelastic), is to first order valid but somewhat 
simplistic. The imbalance of neutrons to 
protons does control the physical mechanisms
of the elastic isovector piece whereas the number of 
absolute neutrons (for proton scattering) or 
the number of absolute protons 
(for neutron scattering) are the source of strength for the 
inelastic component.
Although this type of impulse microscopic
approximation does not intrinsically contain multiple scattering, correlations,
exchange, or coupled channels, it is a good approximation at higher energies 
(it has been shown to work adequately at 
projectile energies of 150 MeV~\cite{amos1}) and it signifies that
beyond first order the asymmetry term needs a theoretical re-evaluation. 

Writing the asymmetric potential
in macroscopic nucleon-nucleus optical potential form is also illuminating: 
\begin{eqnarray}
V_{asym} = V_{isoNA}(\tau_{proj}\cdot\tau_{targ}) \nonumber \\
= V_{isoNA}(\tau_{proj_+}\tau_{targ_-}+\tau_{proj_-}\tau_{targ_+}+
\tau_{proj_z}\tau_{targ_z}), \nonumber \\
\end{eqnarray}
and likewise the difference equation subtracting neutron scattering from
proton scattering is
\begin{eqnarray}
V_{asym}(t_{proj}=+\frac{1}{2})-V_{asym}(t_{proj}=-\frac{1}{2}) \nonumber \\
\approx V_{isoNA}(\tau_{proj_-}\tau_{targ_+}-\tau_{proj_+}\tau_{targ_-}+
\tau_{targ_z}). \nonumber \\
\end{eqnarray}
This form again demonstrates a direct connection to both elastic scattering and
inelastic charge-exchange. The elastic $z$ component piece is proportional
to $\tau_{targ_z}=N-Z$, which confirms the earlier microscopic results.
Similarly 
the inelastic piece is again ambiguous. In many 
nuclei the ground state has an isospin vector designation such that  
$\tau_{targ_z}$ is the maximum value and thus 
$|\tau_{targ_-}|=|\tau_{targ_+}|\propto\sqrt{N-Z}$ but with unstable 
deformed nuclei this is not always the case~\cite{isospinmixing} and 
therefore a general rule about the strength of the
inelastic piece should be treated with apprehension, especially
at higher energy~\cite{satchler2,bauge1}. Both 
the microscopic and macroscopic formulations of the optical potential lead
to a differentiation  between the elastic and inelastic charge-exchange component's
constant of proportionality as seen in previous 
work~\cite{satchler2,lane1,satchler1,carlson1,grimes1,bauge1}.  

All the global optical potentials, 
perhaps because of the linear $N-Z$ term, have trouble
giving excellent results for the calcium and chromium isotopes 
(Fig.~\ref{pdcs3} and Fig.~\ref{pdcs4}). Similarly, as exploration
further from
the line of stability occurs, it is realized that using a stark absolute linear term
leads to erroneous results. For example using a 100 MeV 
neutron projectile gives physical
results when scattering from $^{12,13,14,15}$C using the WP OMP
but it begins to give negative total 
cross sections for the same scattering observables if the target is
$^{16}$C. Obviously all elements will return
non-physical results when the asymmetric term allows for
unimpeded linear growth as the neutron-proton imbalance gets larger (especially
if this term is large as is the case for carbon for the WP OMP).
In general the linear $N-Z$ structure
is a good first approximation 
but a better formulation should be developed for exotic nuclei optical potentials
far from the line of stability.

\subsection{Lane Analysis}\label{subsec:lane}
A Lane consistent 
potential~\cite{lane1,globalB,carlson1,globalJ,grimes1,bauge1}  
guarantees 
a near equivalent isoscalar and isovector nuclear 
potential for proton and neutron projectiles 
at the same initial projectile bombarding energies and thus the measure 
of Lane consistency is a good
check on the integrity of the isovector term.

As discussed in Sec.~\ref{subsec:isoscale}, the
volume potential of all traditional OMPs
can be split into the isoscalar and
isovector parts respectively
\begin{equation}
{\cal V}_V(E)= V_0(E)+{\cal I}(E),
\end{equation}
where ${\cal I}(E)$, the isovector component, has its sign dependent on the isospin
of the projectile.
Because the effective short range projectile kinetic energy
of the proton-nucleus potential is different than in the corresponding
neutron-nucleus potential, a Lane consistent potential
will add a coulomb correction term,$\Delta_c$ to
the traditional proton-nucleus OMP,
\begin{equation}
{\cal V}_V(E)= V_0(E)+{\cal I}(E)+\Delta_c,\label{cc_0}
\end{equation}
which
adjusts the energy dependent proton-nucleus potential to account for the 
lessening of the initial bombarding energy of the 
charged projectile as it heads towards the target due to the long range 
coulomb field~\cite{lane1}. To be
all inclusive this correction
should be included for the complex volume and
spin orbit pieces, traditionally however it
has only been applied to the real central term (recent exceptions are
Ref.~\cite{bauge1,globalMR}). 

Using the notation of Eqs.~\ref{start}-\ref{finish}
of this WP potential the real central coulomb correction, $\Delta_c$, can easily
be derived as
\begin{equation}
{\cal V}_V(E)+{\cal V}_S(E)
={\cal V}_V(E-f_{coul})+{\cal V}_S(E,-f_{coul})+\Delta_c(E),
\end{equation}
where $E$ is the original projectile bombarding energy 
and the volume,surface and coulomb potentials are defined
using Eqs.~\ref{coul}-\ref{coulomb},~\ref{start}-\ref{eq3},\ref{eq7}.
This equation leads directly to a solution as
\begin{equation}
\Delta_c = {\cal V}_V(E)+{\cal V}_S(E)-{\cal V}_V(E-f_{coul})
-{\cal V}_S(E-f_{coul}),\label{cc_exact}
\end{equation}
which is effectively the difference between the original and an
energy adjusted  central volume term. All three of the OMPs examined
contain a coulomb correction term.

\begin{figure}[ht]
\includegraphics*[width=3.35in,angle=0]{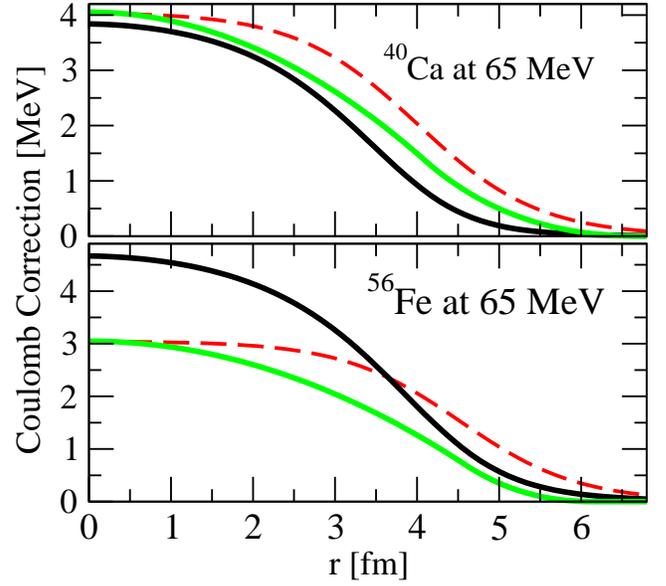}
\caption{(Color online) These are two example depictions at
65 MeV projectile energy of the functional
coulomb correction potential for the targets $^{40}$Ca and $^{56}$Fe respectively.
The solid black line is the magnitude of the exact
calculation (the coulomb correction is actually 
negative), for the potential of this work, using Eq.~\ref{cc_exact}.
If the coulomb correction term is applied to the proton-nucleus
optical potential
then the potential can be considered isovector Lane consistent (described
in the text).
The solid green line is the approximate function which was determined
in this work by fitting the global nucleon-nucleus data sets. This
approximate form was determined by Eq.~\ref{cc2}. The thinner red dashed line
is given as an example Woods-Saxon functional form for comparison.
}
\label{coul_func}
\end{figure}
Equation~\ref{cc_exact} represents
the exact definition of the real coulomb correction. 
Historically, starting with Ref.~\cite{globalB}, an ansatz
was made to 
use a Woods-Saxon potential form with the strength proportional to the average
coulomb potential
in the short range to approximate this coulomb correction, 
both the KD
and MD OMP take this approach. The WP OMP uses a different approximate form,
instead of attaching the coulomb correction term to the volume 
term (with Woods-Saxon shape) it is combined with 
the coulomb term by having the
radius of the short term coulomb potential, Eq.~\ref{eq22}, becomes 
artificially larger and  uncharacteristically 
energy and atomic number dependent. The 
coulomb correction of this work is defined as
\begin{equation}
\Delta_c \approx\left (f_{coul}({\cal R}_{C})-f_{coul}({\cal R}_{C_0}
=1.20\times A^{\frac{1}{3}} fm)\right )\label{cc2},
\end{equation}
in which a literal addition to the coulomb potential was devised by extending 
${\cal R}_{C}$ beyond its traditional value. To extract this coulomb correction 
to the nuclear term
we subtracted the known short ranged coulomb potential from the 
full-extended fitted coulomb potential of this work as Eq.~\ref{cc2} details.

This non-traditional approach has some advantages.
First it 
disentangles the unknown coulomb correction from the likewise unknown nuclear
volume term and combines it to the more apparent short range 
coulomb force. Second, as developed in Ref.~\cite{globalJ}, 
a Woods-Saxon shape is not an excellent functional 
representation of the exact coulomb correction 
result (see Fig.~11 of Ref.~\cite{globalJ} for example). 
The approximate coulomb
correction of this work (Eq.~\ref{cc2}) more satisfactorily
represents the shape of the
exact result. 
Giving an example, in Fig.~\ref{coul_func} 
are depicted two examples at the commonly tested experimental
projectile energy of 65 MeV. The black line represents the calculation of the 
exact result, the green line is the function which 
this research developed in the process
of fitting the coulomb radius to the global 
data set using Eq.~\ref{cc2}. The thin dashed
red line is the Woods-Saxon shape normalized to this 
approximation added as a reference. Although the green fit of this research does
not always match the exact result, its 
functional shape is closer to
the exact result 
specially in the important interior region. Another reason to use this
approximation is that
this correction has an operational advantage that it can be
easy be added to modern optical codes like ECIS~\cite{ecis} by changing the
value of the coulomb radius.
The exact result of Eq.~\ref{cc_exact}, which does not have a 
closed analytical form, cannot be added neatly.

\begin{figure}[ht]
\includegraphics*[width=3.35in,angle=0]{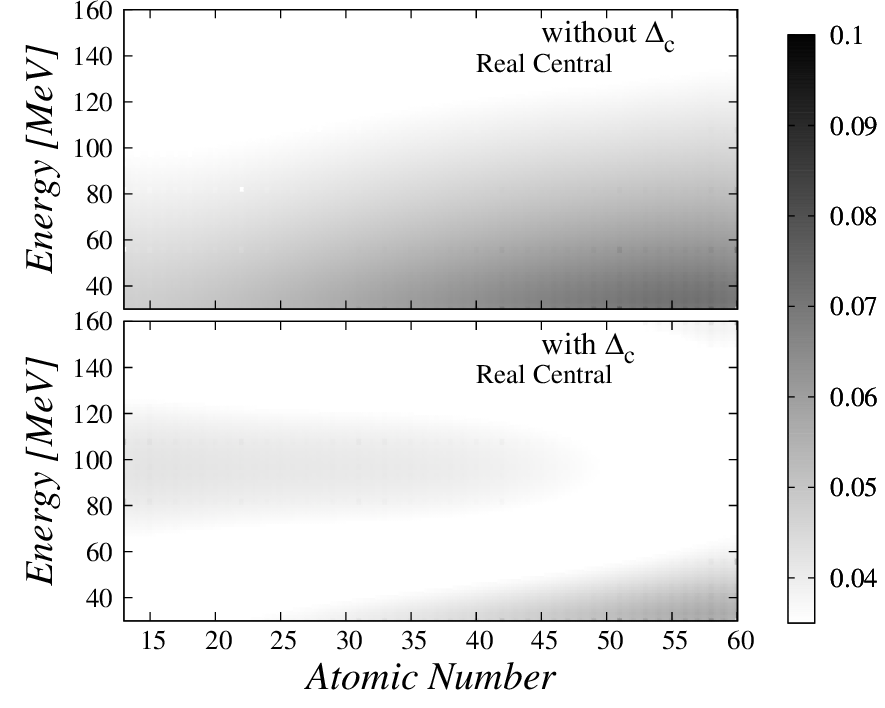}
\caption{This is a depiction in projectile energy-target atomic number phase space
of the {\bf real} fraction Lane inconsistency relative to the magnitude of 
the total moduli of the central potential for sample nuclei. The top panel shows the 
optical potential of this work 
{\bf without} a coulomb correction 
term assuming only ${\cal R}_c=1.20 \times A^{\frac{1}{3}}$ fm.
The bottom panel depicts the same but 
{\bf with} a coulomb correction term assuming a dynamic $E$ and $A$ 
dependent coulomb radius,
${\cal R}$, described in the text.
The unweighted average for the whole phase space of the top panel is $0.043$ Lane
inconsistency.
The unweighted average for 
the whole phase space of the bottom panel is $0.034$ Lane
inconsistency. The fraction of $0.04$ is assumed here to be a pragmatic minimum and a
measure of sufficient
Lane isovector consistency.
The sample nuclei for each atomic number were near or on the line 
of stability and are the same as those in Fig.~\ref{isoscale}.}
\label{contour1}
\end{figure}
\begin{figure}[ht]
\includegraphics*[width=3.35in,angle=0]{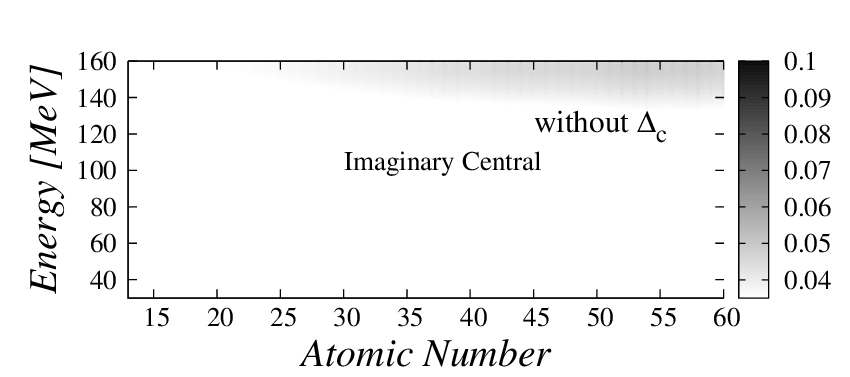}
\caption{
This is a depiction in projectile energy-target atomic number phase space
of the {\bf imaginary} fraction Lane inconsistency relative to the magnitude of
the total complex central force for sample nuclei which are listed in
Fig.~\ref{isoscale}. Overall the isospin
inconsistencies are smaller than its real counterpart until $E>140$ MeV.
The unweighted average for the whole phase space is $0.025$ Lane
inconsistency. The fraction of $0.04$ is assumed here to be a pragmatic minimum and a
measure of sufficient
Lane isovector consistency.
}
\label{contour2}
\end{figure}

To examine if this inclusion of a coulomb correction term brings the
WP OMP closer to Lane
consistency the contour graphs of Figs.~\ref{contour1}-\ref{contour2} are 
introduced. They
map over the complete applicable projectile energy and 
atomic number of target phase space for the WP OMP
and the measure of Lane inconsistency is normalized to 
the moduli of the central potential. The
figures have ranges between 4\%  
and 10\% Lane inconsistency. 
It is noted that of reactions in this phase space the real 
central coulomb correction is at
worst about a  12\% correction. Fig~\ref{contour1}, top panel, 
shows the level of inconsistency without a 
coulomb correction (with an average inconsistency of 4.3\%). 
The systematic need for a coulomb correction for the present 
potential is apparent especially
at lower energies and larger targets, these sensitivities are congruent 
with the coulomb correction term developed
in the KD OMP of Ref.~\cite{globalK} which increases linearly with $Z$ and to good
approximation decreases linearly with $E$.

The real central coulomb correction is inherently included in 
this potential, as discussed above,
and with it the WP potential is substantially more Lane consistent as depicted in
Fig.~\ref{contour1}, bottom panel. 
Much more of the phase space is now below the 4\% level and
only in a few small regions does the coulomb correction addition make the potential
less Lane consistent (high $A$ and high $E$ and low $A$ at midrange energies). The
areas where the coulomb correction was substandard is where the 
experimental data is sparse, especially lacking neutron reactions. 
For example, two experimentally popular 
projectile bombarding energies are at 65 MeV and 135 MeV. Here, with the 
coulomb correction included 
(as shown in the bottom panel of Fig.~\ref{contour1}), the Lane consistency
is very good. The
energies between these guideposts are where the fit has difficulties. The
average value of Lane consistency with the coulomb correction is 3.4\%. In our studies
developing this analysis it was found that using a constant energy independent
${\cal R}_c$ between $(1.1 \times A^{\frac{1}{3}})$ fm and 
$(1.35 \times A^{\frac{1}{3}})$ fm to represent the fixed traditional short
range coulomb radius
in Eq.~\ref{cc2}
produced the best results which 
is reassuring because it matches the standard short range
coulomb potential found in the literature.

The imaginary coulomb correction term has been ignored in the KD, MD, and this
potential.
It has been argued that it can
be neglected because it is small~\cite{rapp1,grimes1} or 
because of lack of sufficient 
experimental data
it will be ambiguous~\cite{globalK}. 
In performing this analysis we found it to be small at
low energy but unfortunately at high 
energy (greater than 120 MeV projectile energy) it
begins to be significant in contrast with the real coulomb
correction term which diminishes as 
depicted in Fig.~\ref{contour2}. 
The average of the Lane inconsistency over
the entire phase space normalized to the magnitude of the complex WP central potential
measures 3.4\% in the real term and 2.5\% for 
the imaginary term. 
However the imaginary Lane 
inconsistency can run higher than 9\% for 
$A>60$ and $E>160$ MeV. 

Examining the 
quality of the 
results in reproducing experimental reactions in Sec.~\ref{sec:results} there is 
significant error in the neutron total cross section calculation
at high atomic number and high energy as represented in Fig.~\ref{ncs4}. Since this 
potential fits the proton and neutron observables simultaneously this could
be due to the lack of that imaginary coulomb correction term. 
In future high energy optical potential work this imaginary correction should
not be neglected. 

The requirement that the optical potential of this work
be completely Lane consistent is not congruent with the goals of this research or
the general character of the nuclear force. 
Charge-dependent 
(isospin symmetry breaking) discoveries~\cite{epelbaum1,miller1}, 
the uncertainty in the short range coulomb potential, and 
the approximate functional forms applied in the phenemological
optical potentials all diminish the importance of attaining
perfect isospin
nuclear symmetry and a Lane potential. 
The goal of under 4\% inconsistency seems to be a reasonable
pragmatic minimum giving these conditions.
This research aimed to be as pure
a phenemological fit as possible letting the reproduction of experimental data
confirm what
microscopic theory has shown elsewhere regarding the
coulomb correction~\cite{globalJ,bauge1,Li1}. The conclusion is that there is
good agreement
when there is a copious amount of
experimental data giving credence to the technique employed.

\subsection{A comparison}\label{subsec:compare}
A comparison of the three global optical potentials isovector asymmetry terms,
which use the traditional $N-Z$ terms, now follows. Here 
dramatic differences between the three formulations can be isolated.

Starting from the simplest potential, the MD OMP
has a neutron excess asymmetric term for the real volume 
and the real spin-orbit amplitude only:~\cite{globalM}
\begin{eqnarray}
{\cal I}_{MD} = \pm \frac{N-Z}{A}16.5 f_{WS}(r,{\cal R}_i,{\cal A}_i)
\nonumber \\
\mp \frac{N-Z}{A} 3.75 
\frac{d}{dr}f_{WS}(r,{\cal R}_{SO},{\cal A}_{SO})({\bf l\cdot\sigma}),
\label{as:md}
\end{eqnarray}
which has no explicit energy dependence and has the standard 
linear term of $\frac{N-Z}{A}$. The difference between the 
proton projectile and the neutron projectile is simply a sign
change, the internal
geometry parameters have no isospin dependence. 

The KD OMP has an explicit asymmetry term 
only for the real volume component and the imaginary surface component~\cite{globalK}
\begin{widetext}
\begin{eqnarray}
{\cal I}_{KD}=\pm \frac{N-Z}{A}\big(21.0 (1-v_2(E-E_f)+v_3(E-E_f)^2-7.0\times 10^{-9}
(E-E_f)^3)  f_{WS}(r,{\cal R}_{iv},{\cal A}_{iv})\big)
\label{as:kd} \nonumber \\
\mp i4{\cal A}_{d}\;\frac{N-Z}{A}\big(16\frac{{(E-E_f)}^2}{{(E-E_f)}^2+{d_3}^2}
\exp {(-d_2(E-E_f))}\frac{d}{dr}(f_{WS}(r,{\cal R}_{d},{\cal A}_{d})\big),
\end{eqnarray}
\end{widetext}
where $v_2,v_3,d_2$ and $d_3$ are functions which depend on the nucleon number of
the target, and projectile energy and the isospin character of the projectile. 
Likewise $E_f$ 
represents the Fermi energy of the target, extracted from mass excess
values~\cite{Audi2}, and is dependent on the isospin of the projectile. The
internal
geometry parameters have no explicit isospin dependence.
Because of the separate functional dependence on the projectile 
the asymmetry term,$I_{KD}$ of Eq.~\ref{as:kd}, is not
exactly linear and the isospin flip in the projectile is not simply a sign
change as with the $I_{MD}$ term given by Eq.~\ref{as:md}.

The optical potential of this work (WP OMP), 
outlined in Sec.~\ref{subsec:ours}, has vector isospin asymmetry $(N-Z)$
in five
major terms: real and imaginary volume, imaginary surface, and real and
imaginary spin orbit (Eqs.~\ref{eq2},\ref{eq5},\ref{eq9},\ref{eq11},\ref{eq13}): 
\begin{eqnarray}
{\cal I}_{WP}=\pm (N-Z) (V_{V_i} +  i W_{V_i}) f_{WS}(r,{\cal R}_i,{\cal A}_i)
\nonumber \\
\mp i4 (N-Z) {\cal A}_S W_{S_i} \frac{d}{dr}f_{WS}(r,{\cal R}_S,{\cal A}_S) 
\nonumber \\
\mp (N-Z) (V_{{SO}_i} +  i W_{{SO}_i})
\frac{d}{dr}f_{WS}(r,{\cal R}_{SO},{\cal A}_{SO})
\end{eqnarray}
where $V_{V_i}, W_{V_i}, W_{S_i},V_{{SO}_i},$ and $W_{{SO}_i}$ are
separable polynomial functions in terms of projectile energy and
nucleon number (Eqs.~\ref{start}-\ref{finish}). 
There is no projectile isospin dependence
within the polynomials. This potential also enforces the separability 
of $E$ and $A$ by using 
$(N-Z)$ and not $\frac{N-Z}{A}$. There are some explicit asymmetry terms in the
${\cal A}_V$ (Eq.~\ref{eq16}) and ${\cal R}_{SO}$ (Eq.~\ref{eq20}) 
geometry terms which also lead to
non-symmetric neutron excess terms. This optical
potential has therefore attempted
to fit the imaginary volume, surface, and spin orbit asymmetry 
terms, the other contemporary global OMPs discussed within have
set their explicit imaginary asymmetric terms to zero.

To explore the differences between the three 
OMPs further let us again examine volume integrals
which have been illustrative in the past to help describe Gamow-Teller 
and Fermi charge-exchange transitions~\cite{osterfeld}
and to effectively compare the strength of disparate shaped optical potentials.
Explicitly calculated are, using the notation of Eq.~\ref{WS}, the following integrals:
\begin{widetext}
\begin{eqnarray}
J_{V_V}/A = &&-\frac{4\pi}{A}\int_0^\infty r^2 \Big({\cal V}_V(E,A,N,Z,{\cal P},MN)\Big)
f_{WS}(r,{\cal R}_V,{\cal A}_V)\; dr 
\label{volume1} \\
J_{W_V}/A = &&-\frac{4\pi}{A} \int_0^\infty r^2 \Big({\cal W}_V(E,A,N,Z,{\cal P},MN)\Big)
f_{WS}(r,{\cal R}_V,{\cal A}_V)\;dr  
\label{volume2} \\
J_{W_{S}}/A = && +\frac{4\pi}{A}4{\cal A}_S\int_0^\infty r^2
\Big({\cal W}_D(E,A,N,Z,{\cal P})\Big) 
\frac{d}{dr}f_{WS}(r,{\cal R}_S,{\cal A}_S)\;dr.
\label{volume3}
\end{eqnarray}
\end{widetext}
These equations will be used to calculate the differences in volume
integrals between a proton projectile and a neutron projectile acting upon
the same target nucleus (the isospin asymmetry).
The difference is emphasized, as it was in the linear study of the $N-Z$ term,
because this is what characterizes the
isovector asymmetry term from the rest of the dominating isoscalar
nuclear potential; it does not disappear upon subtraction of this 
isospin flip of the projectile. 
In the KD potential, 
where there are two separate potentials with different functions
to differentiate proton and neutron scattering,
this definition of asymmetry is somewhat ambiguous, it is much more 
enhanced than simply the $N-Z$
term. So in this comparison we will define isovector to be half the difference
between the proton-nucleus and neutron-nucleus potential. This will include the 
addition of the nuclear coulomb correction term in the proton potential for all three
of the OMPs. This seems to be the best workable definition for isovector for these
three potentials especially in tandem comparisons with the microscopic potential that
follow.
\begin{figure}[ht]
\includegraphics*[width=3.35in,angle=0]{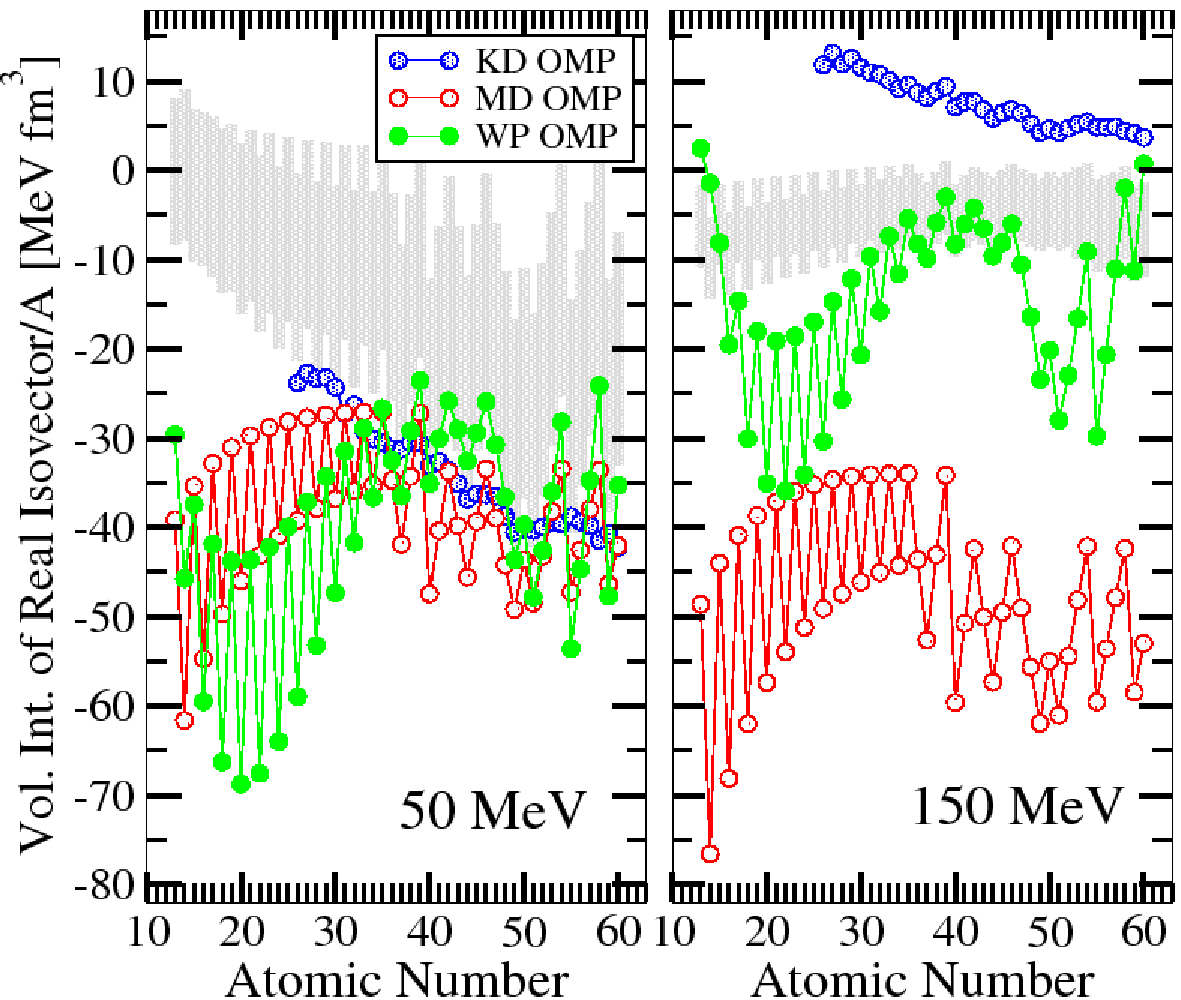}
\caption{(Color online)  This is a calculation of the 
difference in the real component of the
nuclear central potential integrals (volume and coulomb correction) 
between a proton and neutron projectile (about twice the actual
isovector volume integral)
for a representative selection
of nuclei using three different global optical potentials. 
The nuclei were chosen to have nonzero $N-Z$ terms and to
be close or on the line of stability. The nuclei are the same 
as used in Fig.~\ref{isoscale}.
The dark blue semi-filled circular 
points are the calculation using the KD optical potential~\cite{globalK},
the red unfilled circular 
points use the MD optical potential~\cite{globalM} and the light filled green circular
data points
are calculated using the WP potential of this work. The lines connecting the data
points are there to make it easier for the eye to follow. The left panel used a 50 MeV
nucleon projectile and the right panel uses a 150 MeV projectile where the
differences in the three calculations become more pronounced. The light gray
shaded area is the range of values for microscopic potentials which
are described in the text in Sec.~\ref{subsec:isoscale}.
}
\label{assym2}
\end{figure}

In Fig.~\ref{assym2} the two panels contain results of the real
volume, surface, and coulomb correction integral difference 
at 50 MeV (left panel) and 150 MeV (right panel)
energy between a proton and a neutron projectile for the three global optical
potentials (approximately twice the isovector strength). This volume integral
difference is a direct measure of the change in strength that
the three potentials have for the dynamics of a isospin
flip of the projectile on the same $N-Z\ne 0$ target.  
For every nucleon number ($13 \le  A \le 60$) there was chosen a
representative target that is stable or close to the line of stability which
are listed for Fig.~\ref{isoscale}.
The calculation is explicitly of
\begin{equation}
(Re\;J/A)_{iso} = (J_{V_V}/A)_{proton}-(J_{V_V}/A)_{neutron}+J_{\Delta_c}
\label{volume4}, 
\end{equation} 
using the Eq.~\ref{volume1} and the integral of the 
coulomb correction term,$J_{Delta_c}$,
which for the WP OMP has the analytical form
\begin{equation}
J_{\Delta_c}=-\frac{2}{5}\pi\hbar c e Z
\left (({{\cal R}_{C}})^2-{(1.20\times A^{\frac{1}{3}} fm)}^2\right ),
\end{equation}
and for the KD and MD potential the coulomb correction integral
calculation is similar to Eq.~\ref{volume1}
because of its Woods-Saxon functional form.
Figure~\ref{assym2} 
shows some agreement at the low energy (for $28<A<60$) but it is more disparate at
the high energy. Likewise at $A<20$ there
is sharp disagreement between the two applicable potentials (MD and WP) for these
light nuclei. The fine features are telling also, the wildly oscillatory behavior
of the MD and WP integrals are indicative of the size of the neutron excess
(this is to
be expected because they are similar to Gamow-Teller and Fermi sum rules
which are proportional to $N-Z$~\cite{osterfeld}), in the
KD integrals this behavior has been quenched because the neutron and proton potentials
lack the requisite similarity. 

Figure~\ref{assym2} also has a light gray shaded area depicting a range of microscopic 
optical potentials isovector volume integrals. These microscopic potentials 
have been described in Sec.~\ref{subsec:isoscale} and 
their isoscalar volume integrals
depicted in Fig.~\ref{isoscale}. Overall the microscopic optical potentials have a
systematically lower strength than the three OMPs, although it is encouraging
to see that at the lower 50 MeV projectile energy there is a general agreement between
all the potentials for target nucleons with $A>28$. Some features of the microscopic
potential isovector character are that it has a much smaller width at higher energies
because the impulse approximation, where a free density-independent t-matrix
is closer to being realized. Likewise the microscopic potential also shows an 
oscillatory $N-Z$ behavior in the isovector volume strength, for example the 
isovector character of $^{14}$C is almost twice as big as $^{13}$C because it has
twice the neutron excess. At high energies the KD and WP isovector elements shrink
and mimic the microscopic result better than the MD OMP which grows; this
is deceiving, for the source of the magnitude shifts are not 
substantially isovector in origin but are almost entirely due to the
coulomb correction which is very much reduced in the WP and KD optical potentials
at high energies but is
energy independent in the MD optical potential.

\begin{figure}[ht]
\includegraphics*[width=3.35in,angle=0]{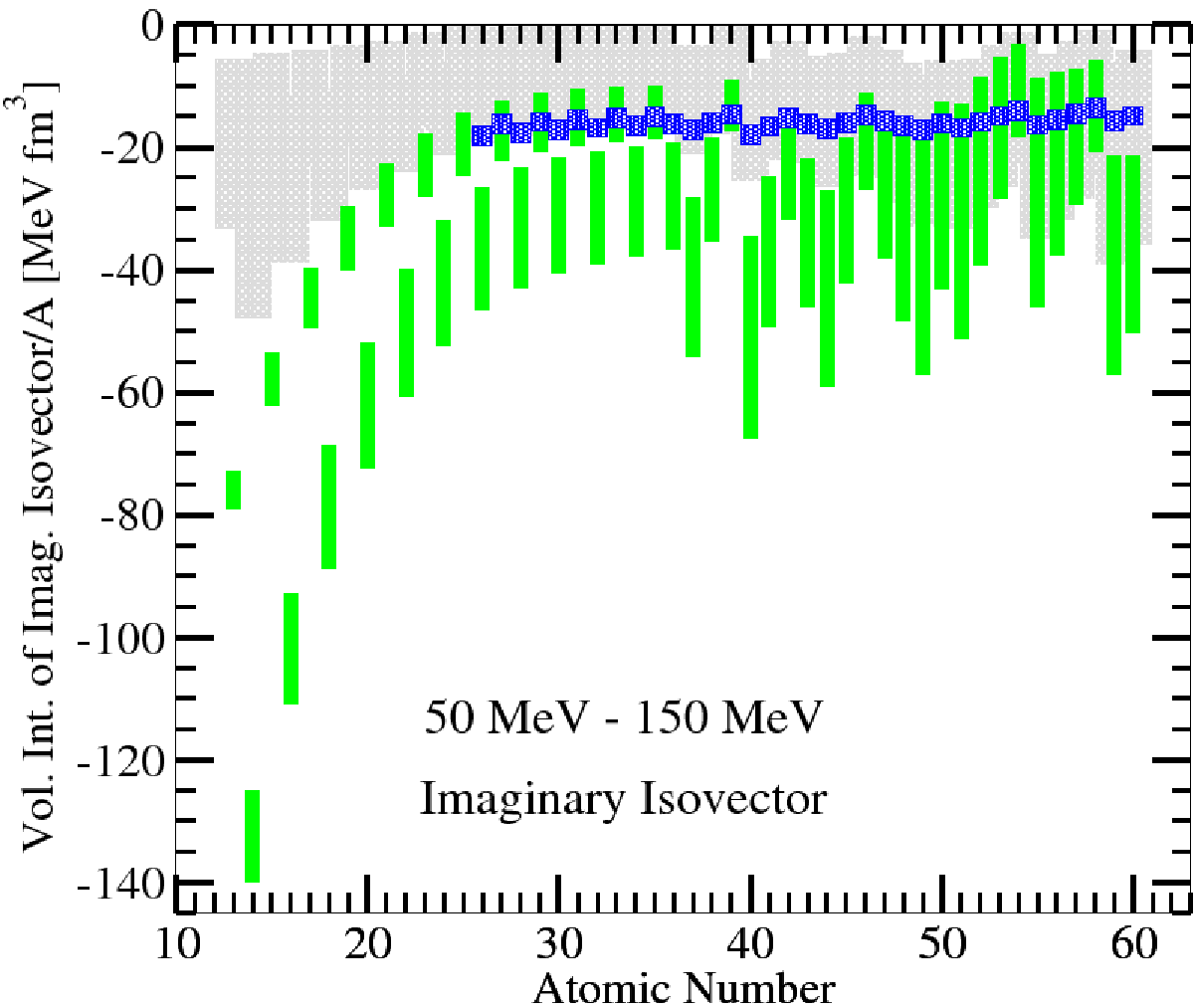}
\caption{(Color online) 
This is a calculation of the difference in the imaginary component of the 
central volume integrals
between a proton and neutron projectile (about twice the isovector strength)
for a representative selection
of nuclei using the WP and KD global optical potential. 
The nuclei were chosen to have nonzero $N-Z$ terms and to
be close or on the line of stability, the specific nuclei are listed
in Fig.~\ref{isoscale}.
The light green fill represents the range possible for nucleon projectile energies
between 50 MeV and 150 MeV 
for the WP global optical potential of this work. The KD global optical
potential (dark blue fill)
is smaller and has a diminished range but its values are comparable
to its real counterpart.
The light gray
shaded area is the range of values for microscopic potentials which
are described in the text in Sec.~\ref{subsec:isoscale}.
}
\label{assym3}
\end{figure}
The calculation
of the explicit 
imaginary volume asymmetric potential is a distinct attribute of the WP 
optical potential described by this work. Calculating the isospin difference
of the imaginary term for this potential is executed by:
\begin{eqnarray}
(Im\;J/A)_{iso}&=&(J_{W_V}/A+J_{W_{S}}/A)_{proton} \nonumber \\
&-& (J_{W_V}/A+J_{W_{S}}/A)_{neutron},
\label{volume5}
\end{eqnarray}
where both the volume and surface terms are included 
as described by Eqs.~\ref{volume1}-\ref{volume3}.
In contrast the KD OMP only has an implicit small imaginary asymmetric surface term
but in subtracting a neutron projectile reaction from a proton projectile reaction
off the same target nucleus there is a significant imaginary difference  coming
from the disparity in isospin dependent volume strengths and Fermi energies.
The MD potential has an imaginary isovector strength of zero. 
In Fig.~\ref{assym3} a plot of these imaginary 
volume integrals are depicted at a range between 
50 MeV and 150 MeV projectile energies. The
results are striking, this imaginary volume integral is quite large, often the 
same size or larger than
the companion real piece. 
Having an 
imaginary asymmetry term, as the WP and KD optic potentials have,
allows for a more nuanced optical potential which contains
mechanisms for charge-exchange 
resonances~\cite{osterfeld}, important at these energies.
Again, for further comparison, the microscopic optical potential range for the 
imaginary isovector element is depicted in light gray.
It is systematically lower than the phenemological optical 
potentials but is also is often much stronger than its real counterpart, especially at
higher energies. The microscopic potential integral 
still carries large remnants of the $N-Z$  character
in the results as does the WP OMP. The KD OMP is greatly reduced reminiscent of the 
real isovector volume element of Fig.~\ref{assym2}. Incidentally, if a realistic
imaginary coulomb correction is added 
it would  give a small positive volume element differential to the results
depicted in Fig.~\ref{assym3} thus leading to better agreement between the 
phenemological and microscopic potentials for this term.

A comparison of these asymmetric isovector volume integrals differences
can me made to experiment. 
It is well understood that the
asymmetry volume integral is tied to Gamow-Teller
transitions in charge-exchange reactions
which dominate at
high energies~\cite{osterfeld}. The volume integral equations of 
Eqs.~\ref{volume1}-\ref{volume5} are equivalent to doing a Fourier transform
to momentum transfer ($q$) space set to zero. In this forward scattering case the 
inelastic aspect of the 
asymmetry terms dominates. This is because in the coulomb distorted wave basis the long
range coulomb potential, which dictates extremely forward angle proton scattering, 
is external to the optical potential so the magnitude of the elastic scattering 
component is zero, likewise
the neutron scattering elastic forward amplitude is zero. This extreme
scattering has been recognized as being important in determining Gamow-Teller
strengths~\cite{goodman1,fujita1}.
If the difference is taken between the energy needed to initiate a proton-neutron
charge-exchange and a neutron-proton charge-exchange for the same target nucleus
(the difference in Q values derived from the mass excess values for the resultant
final nuclei) this can be compared to the difference in the
volume integrals which
give the minimum energy, zero momentum transfer, nuclear density difference.

The volume integral for the coulomb potential is infinite but
it can easily be defined within the traditional nuclear radius,
using the short range expression, used by all three OMPs under
examination, which states that when $r<{\cal R}_C$:
\begin{eqnarray}
J_C/A=\frac{4\pi}{A} \int_0^{{\cal R}_V \;A^{\frac{1}{3}}}f_{coul}(r,{\cal R}_C,A,N,Z)
r^2 dr. \label{volumecoulomb} 
\end{eqnarray}
In the 
short range the coulomb potential interferes with 
the nuclear force and contributes to the 
overall strength of the asymmetric volume element. 
The KD and MD optical potentials also include a separate coulomb correction term
while the potential of this work (WP) does not contain a separate coulomb correction term
but is part of the original short-range coulomb term as detailed in
Sec.~\ref{sec:iso}. These nuclear corrections are significant in 
the short range and must be included when illustrating short range differences between
neutron and proton scattering.

To calculate the difference per nucleon
in potential energy density at zero momentum transfer in the short range
five terms are calculated
\begin{eqnarray}
J_{iso}&=&(Re\;J/A+J_{coul. corr.}/A)_{iso} \nonumber \\
&+& (Im\;J/A)_{iso} \nonumber \\
&+&J_C/A-\Delta J_{sr},
\label{volume_final}
\end{eqnarray}
where the first line
is the real nuclear potential energy density difference
with the coulomb correction included (Fig.~\ref{assym2}). 
The second line is the imaginary nuclear potential
energy density
difference (Fig.~\ref{assym3}.
The last line is the short range coulomb potential volume (Eq.~\ref{volumecoulomb}),
up to the nuclear radius ${\cal R}_V$, 
and a zero point correction, $\Delta J_{sr}$.
This  correction is needed because
of the weakened short ranged 
coulomb volume that is used by all three potentials as
dictated by Eq.~\ref{coulomb}. 
Since the coulomb distorted wave basis sets zero nuclear 
scattering as defined by the traditional 
long range coulomb form (Eq.~\ref{coul}) 
and the short range volume element form actually used
is 20\% weaker
then the traditional coulomb potential, the true zero-scattering point
has been shifted in the distorted wave 
coulomb basis of neutron and proton scattering within the short range 
and this modification has to be normalized accordingly.
Thus the correction is
\begin{equation}
\Delta J_{sr} = (0.2)\frac{4\pi}{A} \int_0^{{\cal R}_C \;A^{\frac{1}{3}}}
 \frac{Ze^2}{r} r^2 dr,
\end{equation}
which is applied to all the potentials consistently.

\begin{figure}[ht]
\includegraphics*[width=3.35in,angle=0]{assym4.eps}
\caption{(Color online) This plot depicts 
the total normalized energy difference ($J_{iso}\rho_{nuc}$)
for the proton and neutron projectiles on a representative selection of nuclei using
two different global optical potentials using potentials at projectile energy of 
150 MeV.
The nuclei were chosen to have nonzero $N-Z$ terms and to
be close or on the line of stability,
the specific nuclei are listed
in Fig.~\ref{isoscale}.
These calculations are compared
to the difference in experimental mass excess energies for the 
charge-exchange reactions, $Q(n,p)-Q(p,n)$ from the same targets, 
the experimental data (dark black circles) are from
Ref.~\cite{Audi}. The MD calculations, the unfilled red circles, 
have multiplied the results of Eq.~\ref{volume_final} by .38 $fm^{-3}$, the WP
calculations, the light filled green circles, 
has been multiplied by .16 $fm^{-3}$. The
experimental mass excess energies shown here were not used to constrain the
OMPs during the fitting procedure. Two other potential calculations along
with experiment are shown in Fig.~\ref{assym5}.} 
\label{assym4}
\end{figure}
\begin{figure}[ht]
\includegraphics*[width=3.35in,angle=0]{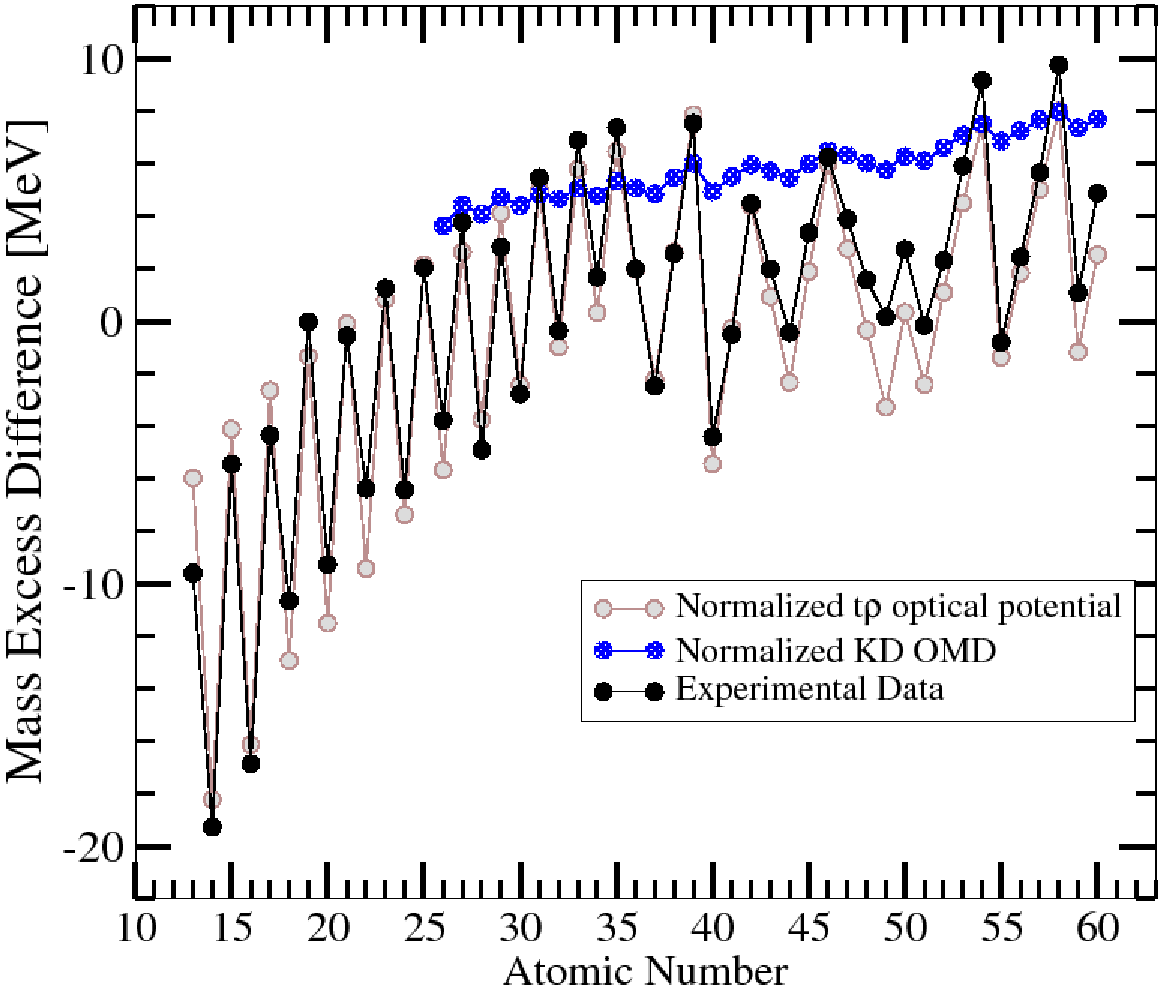}
\caption{(Color online) This plot depicts
the total normalized energy difference ($J_{iso}\rho_{nuc}$)
for the proton and neutron projectiles on a representative selection of nuclei using
two different global optical potentials using potentials at projectile energy of 
150 MeV.
The nuclei were chosen to have nonzero $N-Z$ terms and to
be close or on the line of stability,
the specific nuclei are listed
in Fig.~\ref{isoscale}.
These calculations are compared
to the difference in experimental mass excess energies for the
charge-exchange reactions, $Q(n,p)-Q(p,n)$ from the same targets,
the experimental data (dark black circles) are from
Ref.~\cite{Audi}. The KD calculations, the filled blue circles,
have multiplied the results of Eq.~\ref{volume_final} by .17 $fm^{-3}$, 
the lighter gray circles (following the technique described in Ref.~\cite{med2}),
has been multiplied by 1.14 $fm^{-3}$. The
experimental mass excess energies shown here were not used to constrain the
OMPs during the fitting procedure.}
\label{assym5}
\end{figure}

In Fig.~\ref{assym4} and Fig.~\ref{assym5} there are depicted 
Q value (mass excess) differences for charge-exchange reactions ($Q(n,p)-Q(p,n)$)
in black
along with normalized 
volume integrals for the MD, WP, KD, and a typical microscopic $t-\rho$
optical potentials~\cite{med2} for the forty-eight different sample
target nuclei at a 
projectile energy of 150 MeV. The normalized optical potentials were calculated
by using Eq.~\ref{volume_final} and then multiplying that result by a constant density. 
The functional form results of the optical potential volume integrals 
are in remarkable agreement with the 
experimental $Q$ value differences. Three of the four optical potentials mimic the
shape and structure of the experimental mass excesses, following the global
minimum at $^{13}$C, local maxima and minima at $^{39}K$ and $^{40}$K (dictated
by factors of $N-Z$), 
and the general trends as nucleon number increases.
A natural consequence of fitting elastic scattering data with a proton
and neutron inclusive optical potential is
the ability to mimic the Q values of the charge-exchange reactions at high energies
given a
simplistic constant 
density normalization factor even without this data being used as a fitting constraint.
This structure is remarkably also built into the the microscopic optical potential
which is a sum of the two-body nucleon-nucleon potentials~\cite{med2}. The
only optical potential that does poorly is the KD potential which was developed
using separate potentials for proton and neutron scattering. The general conclusion
is that by simultaneously fitting neutron and proton projectiles (or neutron-proton
and proton-proton phase shifts in the case of the microscopic potentials) one
automatically develops a potential which has the correct target dependant
relative moduli strengths for the
forward scattering volume integrals at high energy which are tied directly to
both the mass difference and Gamow-Teller matrix elements. 

\begin{table*}
\begin{tabular}{|c||l||c|c||c|c|c|}
 \hline
Energy& Model & $\Delta Re J$  & $\Delta Im J$ &
$\Delta \sigma_{elast}$ & $\Delta \sigma_{react}$ &$\Delta \sigma_{tot}$\\
MeV &&  MeV fm$^3$ & MeV fm$^3$ & mb & mb & mb \\
 \hline
&KD & +1.8   &   +0.4    &  +14 &  -28 & -14 \\
50 &MD & +14.5  & 0.0    &  -36  & -9  & -45 \\
MeV&WP & -10.8   &  +8.2       &  +68  &  -51 & +17 \\
&$t-\rho$&+14.0&  +7.5    &  -18   &  -22  &-40 \\
\hline
&KD & +2.6 & +0.5    & -19  & -4   & -23 \\
150&MD & +18.2& 0.0    & -106 & -5   & -111 \\
MeV&WP & +3.1 & +16.8  & +188 & -287 & -99 \\
&$t-\rho$&+2.1 & +5.8   & -6& -15    &-21 \\
\hline
\end{tabular}
\caption{This is a comparison of the strength of the asymmetry term at $A=40$.
The calculation is the difference between neutron-nucleus total cross section
calculations from scattering off of $^{40}$Ar and
$^{40}$Ca  at 50 MeV and 150 MeV projectile 
energies $(\sigma_{Ar}-\sigma_{Ca})$. The differences between the three optical
potentials at high energies are mostly due to the isovector asymmetric term
and pronounced and easily separable. At lower energies the differences fall within
experimental deviation.
}
\label{T4}
\end{table*}

The general systemic target shape is correct in Figs.~\ref{assym4},\ref{assym5} 
for the MD, WP, and $t-\rho$ potentials but the magnitude differences are substantial.
The constant density factor is 1.14 fm$^{-3}$ for the microscopic
potential, .38 fm$^{-3}$ for the MD 
calculation (the isovector volume strength is three times the size of the 
microscopic calculation) and the WP OMP has .16 fm$^{-3}$ for the constant density 
(the isovector volume strength is seven times the size of the microscopic). 
Additionally the MD optical potential has all its strength in the real component
where the WP and microscopic potentials have isovector character distributed in
both the real and imaginary term.
Others have used 
charge-exchange differential cross-sections to constrain the
nucleon-nucleus 
optical potential asymmetric isovector term. 
Future work on this OMP could take that direction following the procedures 
as outlined in Refs.~\cite{lane2,bauge1} which would help constrain these
isovector magnitudes further.

The ramifications of the 
large asymmetric potential differences can be ascertained also with 
reactions that the global
optical potentials are currently fitted to.
Table~\ref{T4} examines the differences between the $^{40}$Ar(n,*) and
$^{40}$Ca(n,*) cross sections with a 50 MeV and 150 MeV 
neutron projectile (calcium subtracted
from argon). 
The large differences generated in the elastic and inelastic cross section predictions
are highly dependent on the contrasts
in the antisymmetric isovector 
term between the three potentials. This reaction was chosen
because both $A=40$ nuclei are stable and there are dramatic differences between
the potentials at $A=40$ in the asymmetric term.
Using neutrons as projectiles also allows an ignorance of coulomb effects and
coulomb correction terms.
At the 50 MeV projectile 
energy the isovector differences are large however the calculations fall
within one or two standards deviation of experimental error~\cite{winters,finlay2}
At the higher projectile energy of 150 MeV, where it can be assumed
that the born and impulse approximations have validity, the effects 
of correlations, coupled channels and potential distortions are at a minimum.
The extreme dramatic reduction in the inelastic cross section
and increase in the elastic cross section  predicted by the WP OMP differs
wildly with the predicted large decrease in the elastic cross section seen by the MD
potential. Likewise the KD and microscopic potentials 
offers a third very realistic conclusion; that the 
differences between $^{40}$Ar(n,*) and
$^{40}$Ca(n,*) cross sections with a 150 MeV neutron projectile are
relatively minor. There is at present no high energy  $^{40}$Ar(n,*) data to
confer which optical potential is closest to experiment.
Experiments calculating
the $^{40}$Ar cross sections
would be a welcome addition as well as would other high energy neutron 
reaction data
like recent experiments detailed in Refs.~\cite{C12p142ra,reaction2}
which, if use $N\ne Z$ target nuclei, are sensitive to the isovector term.

\section{Conclusion} 
\label{sec:conclusion}
The motivations for this work were to construct a 
phenemological nucleon-nucleus global optical potential that is suited for
a wide range of nuclei targets and projectile energies which are within capacity
of the exotic beam accelerators presently running and under development.
We have succeeded in creating one isospin dependent potential
which fits target nuclei $12\le A \le 60$ and a projectile energy
of $30$ MeV$\le E \le 160$ MeV. It compares well with two other recent global
optical potentials~\cite{globalK,globalM}, its advantages are that it is one 
continuous optical potential which 
is designed to do systematic studies on mirror nuclei
and chains of isotopes (a observable calculator has also been made available
to researchers to quickly use this potential for their own research~\cite{applet}).
We have also included an imaginary
asymmetry term which is missing from other recent global optical
potentials and have given comparative analysis on how the asymmetric potential terms
in the three potentials
dramatically differ from each other, microscopic potentials 
and the ramifications to experimentally testable observables. 
It was also ascertained that a  benefit of fitting the proton and neutron
observables simultaneously is the ability to accurately 
define the structure of the mass excesses for the charge-exchange reactions.

There also was a critical examination of the validity of the 
traditional linear proportionality of $N-Z$ applied in the asymmetry term.
An examination of the calculation of the
calcium and chromium experimental observables produced
at 65 MeV projectile energy show a continuing failure to produce great
fits along the isotopic chain, indicative of a breakdown of this linear 
assymetry anasatz. Further systematic experimental studies of chains
of isotopes would help ascertain the direction of further theoretical studies.

Future work could extend this potential to heavier targets,
test other forms of asymmetry potentials, examine spin-spin terms, 
use charge-exchange information to constrain data and 
determine algorithms to better weight the scattering observables. To
better constrain the terms more elastic scattering data from traditional
and exotic nuclei data is needed especially at the higher energies.

\begin{acknowledgments}
S.P.W. gratefully
acknowledges the hospitality of the Physics Department at The Florida
State University, especially Kirby Kemper, 
during the author's sabbatical leave
where this work was begun.  We are also indebted
to a {\it TeraGrid} grant (PHY060025N), funded by the
{\it National Science Foundation},  in which a bulk of the
computational research was done.
This work was also graciously
supported by Eckerd College internal grants which allowed students
to work on this project over ensuing summers including a special debt
to Nicholas Crotty for retrieving the experimental data
and to Eva Romero Luna for her help in arranging the final figures. 

\end{acknowledgments}
\vfill\eject


\vfill\eject
\end{document}